\documentclass[10pt,a4paper,twocolumn,english,pra,aps,showpacs,floatfix,
groupedaddress,superscriptaddress]{revtex4}

\usepackage{graphicx}
\usepackage[english]{babel}
\usepackage{amsmath}
\usepackage{amssymb}
\usepackage{amsfonts}
\usepackage{bm}
\usepackage{bbm}
\usepackage{nicefrac}

\usepackage{color}
\definecolor{shaded}{gray}{0.4}

\begin{document}
\title{Truncation errors in self-similar continuous unitary transformations}

\author{Nils A. Drescher}
\email{drescher@fkt.physik.tu-dortmund.de}
\author{Tim Fischer}
\email{fischer@fkt.physik.tu-dortmund.de}
\author{G\"{o}tz S. Uhrig}
\email{uhrig@fkt.physik.tu-dortmund.de}
 \affiliation{Lehrstuhl f\"{u}r Theoretische Physik I, Dortmund University of
 Technology, Otto-Hahn Stra\ss{}e 4, 44221 Dortmund, Germany}
\date{\rm\today}
\begin{abstract}
Effects of truncation in self-similar continuous unitary transformations 
(S-CUT) are estimated rigorously. We find a formal description via an 
inhomogeneous flow equation. In this way, we are able to quantify truncation 
errors within the framework of the S-CUT and obtain rigorous error bounds for 
the ground state energy and 
the highest excited level. These bounds can be lowered exploiting symmetries 
of the Hamiltonian. We illustrate our approach with results for a toy model of 
two interacting hard-core bosons and the dimerized $S=\nicefrac{1}{2}$ 
Heisenberg chain.
\end{abstract}


\maketitle

In 1994, Wegner \cite{Wegner1994} introduced the method of 
\emph{continuous unitary transformations} (CUT) to many-body physics. 
Independently, a similar method was developed by G{\l}azek and Wilson denoted 
as \emph{similarity renormalization scheme} \cite{Glazek1993,Glazek1994}
to be used in QED and QCD. In this non-perturbative approach, a Hamiltonian 
is mapped to an effective, renormalized model by a series of infinitesimal 
unitary transformations specified by the flow equation. This effective 
Hamiltonian shows (block-)diagonal structure and can be used to calculate 
ground state and low-energy properties as well as observables 
\cite{Kehrein1997,Kehrein1998}.
A perturbative variant (P-CUT) allows for systematic high-order perturbation 
expansions \cite{uhrig98c,Knetter2000,Knetter2003}.
Additionally, the CUT method provides a generic framework to derive 
effective models for many-particle systems that incorporate the relevant 
physics. The mapping of the Hubbard model to a
generalized $t$-$J$ model illustrates this point exemplarily
\cite{Reischl2004,Reischl2006,Lorscheid2007,Hamerla2009,hamer10}. 

Techniques based on CUT have been applied successfully in various fields of 
many-particle physics as electron-phonon-coupling \cite{Lenz1996,Mielke1997}, 
spin-chains \cite{Schmidt2003,Schmidt2004} and ladders 
\cite{Schmidt2005,Fischer2010}, anyonic excitations 
\cite{Schmidt2008,Dusuel2008,Vidal2008} or non-equilibrium problems 
\cite{Kehrein1997} to mention only a few. For an overview, see Refs.\ 
\onlinecite{Wegner2006} and \onlinecite{Kehrein2006}. 

The basic concept of CUT is to transform the problem of (block-)diagonalising 
a Hamiltonian into to the solution of a differential equation for the 
Hamiltonian
\begin{align}
 \partial_\ell H(\ell) = \left[ \eta(\ell),H(\ell)\right], 
\end{align}
commonly known as flow equation \footnote{Therefore, CUT is often also called 
\emph{flow equation method}.}.
The initial Hamiltonian $H(0)$ is linked continuously to a unitarily 
equivalent (block-)diagonal or otherwise simpler Hamiltonian $H(\infty)$
 in the limit of infinite $\ell$.

A drawback of the CUT method is the creation of more complex interaction in the
 Hamiltonian during the flow. To keep the set of equations finite, it is 
necessary to define a truncation scheme and thereby keep only a finite number 
of multi-particle interactions. These truncation errors result in unknown 
errors for the calculated physical quantities. A common approach is to start 
with a strict truncation scheme, i.e., a scheme comprises only a limited
number of terms in the Hamiltonian,
 and to justify this scheme \emph{a posteriori} by comparing it to calculations
 using less strict truncation schemes \cite{Kehrein1994}.
This yields a pragmatically well-working method to control the quality of the 
calculated quantities.
However, since this approach only compares different approximations, the 
relation to the exact result or rigorous error bounds are out of reach.

In our work, we analyze the effects of truncation to the flow and show how 
the truncation can be treated in a mathematical framework. Thereby, we aim to 
quantify the truncation error \emph{a priori}. This approach 
will lead to rigorous  bounds for the ground state energy and 
the highest energy eigen-value for the  Hamiltonian.

The paper is structured as follows: After this introduction, we give an 
overview of the CUT method. In Sect.\ \ref{struct:math}, we  analyze 
the truncation  and define the truncation error. 
The formalism developed is applied to the 
double hard-core boson model in Sect.\ \ref{struct:toy} to provide
a transparent, simple application. Subsequently, 
we discuss in Sect.\ \ref{struct:extend} the modifications needed for the 
calculation of  truncation errors in extended systems. For further 
illustration,  we show results for  the dimerized Heisenberg chain. 
Finally, a summary is  given.

\section{The CUT method\label{struct:CUT}}

\subsection{Homogeneous flow equation}

Probably, unitary transformations are one of the most widely used techniques 
in studies on Hamiltonians.
They render a description of  a Hamiltonian possible 
in a more appropriate basis in which 
the physical properties can be studied more easily. 
Most desirably, every Hamiltonian 
can be diagonalized by a certain unitary transformation. Unfortunately, this 
transformation is usually unknown.

The basic idea of the CUT-method is not to search for such a transformation in 
one step, but to bring the Hamiltonian successively closer to a simpler shape 
by a series of infinitesimal transformation. Therefore, a continuous flow 
parameter $\ell$ is introduced that parametrizes the continuous unitary 
transformation $U(\ell)$. The Hamiltonian is considered to become a function 
$H(\ell)=U(\ell)H^{(b)}U^\dagger(\ell)$ of this parameter. In this way, 
the initial (bare) Hamiltonian $H^{(b)}$ is linked continuously  by a unitary 
transformation to the renormalized Hamiltonian showing the intended structure 
$H^{(r)}=H(\infty)$ in the limit of infinite $\ell$. By derivation with respect
 to $\ell$, one obtains the flow equation
\begin{subequations}
\begin{align}
\partial_\ell H(\ell)=&\frac{\partial U(\ell)}{\partial \ell}U^\dagger(\ell) 
H(\ell)+H(\ell)U(\ell)\frac{\partial U^\dagger(\ell)}{\partial\ell}
\\
=&\left[\eta(\ell),H(\ell)\right].
\end{align}
\label{eq:flowequation}
\end{subequations}
The antihermitian generator $\eta$ of the transformation reads
\begin{align}
\eta(\ell)=\frac{\partial U(\ell)}{\partial \ell}U^\dagger(\ell)=
-\eta^\dagger(\ell).
\label{eq:generator}
\end{align}
Equation \eqref{eq:flowequation} is linear differential equation for 
the Hamiltonian. We 
emphasize that also all intermediate Hamiltonians $H(\ell)$ conserve the full 
information of the system because they are only written in a different basis.

Since the basis has changed during the flow, observables may not be calculated 
directly using their bare operator $\mathcal{O}^{(b)}$ but have also to be 
transformed by a similar flow equation
\begin{align}
\partial_\ell \mathcal{O}(\ell)=\left[\eta(\ell),\mathcal{O}(\ell)\right]
\label{eq:obsequation}.
\end{align}
The transformation of observables was used first by Kehrein and Mielke  to 
determine correlation functions for dissipative bosonic systems
\cite{Kehrein1997,Kehrein1998}.

\subsection{Generator schemes}

Up to here, the problem of diagonalization has only been 
recast in the form of determining an appropriate generator $\eta(\ell)$. 
The key ingredient of 
the CUT-method is to choose the generator as manifestly antihermitian
operator depending on the flowing Hamiltonian. 
We denote the superoperator $\hat \eta:
H(\ell)\to \eta(\ell)=\hat\eta[H(\ell)]$ as \emph{generator scheme} to
distinguish between the mapping $\hat\eta$ and the function $\eta(\ell)$.
In this way, the flow equation for the Hamiltonian \eqref{eq:flowequation} 
becomes non-linear, while the transformation of observables 
\eqref{eq:obsequation} stays linear.
The generator scheme has to be designed in a way that the flow equation has 
attractive fixed points where the Hamiltonian has the desired structure. 
In this  manner, (block-)diagonality can be obtained by merely integrating 
the flow  equation \cite{Wegner1994,Mielke1998,Knetter2000,Dusuel2004}.

For the first generator scheme introduced by Weg\-ner \cite{Wegner1994},
the Hamiltonian
$H(\ell)=H_\text{d}(\ell)+H_{\text{nd}}(\ell)$ has to be decomposed into a 
diagonal $H_\text{d}$ and a non-diagonal part $H_{\text{nd}}$. The generator 
is defined as a commutator
\begin{align}
\eta(\ell)=\widehat \eta_\text{W}[H(\ell)]
=\left[H(\ell),H_{\text{nd}}(\ell)\right]=
\left[H_\text{d}(\ell),H_{\text{nd}}(\ell)\right]
\end{align}
of the diagonal and non-diagonal-part of the Hamiltonian.
One directly realizes that a vanishing non-diagonality yields a fixed point of
the flow. The proof of convergence for unapproximated systems was given by 
Wegner \cite{Wegner1994} for finite
matrices and extended to infinite systems by Dusuel and
Uhrig \cite{Dusuel2004}. The generator decouples eigen-subspaces of different
energy eigen-values, but it is not able to treat degeneracies. In his original 
work
concerning the $n$-orbital model \cite{Wegner1994}, Wegner noticed
divergences. He could avoid them via taking only terms violating the number of 
quasiparticles into account in the definition of $H_{\text{nd}}$ aiming at 
block-diagonality instead of diagonality. In this manner, the complexity of
the a problem can still be reduced significantly because 
different quasiparticle spaces can be studied separately.
\label{pos:cut-mku}

To overcome the problem of residual off-diagonality due to degeneracies,
Mielke \cite{Mielke1998} introduced a generator scheme on the 
matrix level based on a sign function of index differences 
$\eta_{ij}=\text{sign}(i-j)h_{ij}$ that always yields a diagonal 
Hamiltonian. Independently, Knetter and Uhrig \cite{uhrig98c,Knetter2000}
developed a similar scheme which concentrates more generally on a quasiparticle
 picture. In their approach, the Hamiltonian $H(\ell)=\sum_{ij}H^i_j(\ell)$ is 
decomposed into different blocks $H^i_j$ of terms with 
respect to the number $i$ of
quasiparticles created and the number $j$ of quasiparticles annihilated by the
term. In this notation, the generator scheme acts as
\begin{align}
\widehat \eta_{\text{pc}}[H(\ell)]=\sum\limits_{i,j} \text{sgn}\left(i-j\right)
H^i_j(\ell).
\end{align}
In the limit of infinite $\ell$, the Hamiltonian converges to a
block-diagonal, quasiparticle conserving structure if the spectrum is bounded
from below \cite{Mielke1998,Knetter2000,Dusuel2004}. Blocks with $i\neq j$
decay exponentially with rising $\ell$. During the flow, the quasiparticle
spaces are ordered ascending to their energy
eigen-value\cite{Mielke1998,Heidbrink2002}. This implies that the vacuum state,
i.e., the state with $j=0$, is mapped to the ground state of the 
Hamiltonian if it is not degenerated. 

A special feature of this generator scheme is that it strictly conserves the 
block-band structure of the Hamiltonian during the flow.
A similar generator was used by Stein \cite{Stein1997,Stein1998} in a case
where the sign function was not necessary. In contrast to Wegner's generator 
scheme, the right-hand side of the flow equation for the Hamiltonian 
\eqref{eq:flowequation} is 
only quadratic in the Hamiltonian's coefficients instead of cubic.

A recent development in the field of generator schemes is the ground state
generator \cite{Fischer2010}
\begin{align}
\widehat \eta_{\text{gs}}[H(\ell)]=H^i_0(\ell)-H^0_j(\ell).\label{eq:cut-gs}
\end{align}
The definition resembles the particle conserving scheme, but  it is designed 
to decouple only the zero quasiparticle subspace of a 
system, i.e., the ground state if not degenerated.
It was introduced by Fischer, Duffe, and Uhrig to describe quasiparticles 
decays, since the picture of conserved
renormalized quasiparticles becomes very cumbersome.

Compared to the
particle conserving scheme, $\widehat\eta_{\text{gs}}$ exhibits an enhanced
numerical stability and saves computational ressources due to its fast
convergence. As a drawback, it does not conserve the block-band structure as
$\widehat\eta_\text{pc}$ does. In this work, we will make use of 
both generator schemes, $\widehat\eta_\text{pc}$ and  $\widehat\eta_\text{gs}$.

\subsection{Truncation scheme \label{struct:cut-truncscheme}}

\begin{figure*}
\includegraphics[width=\textwidth]{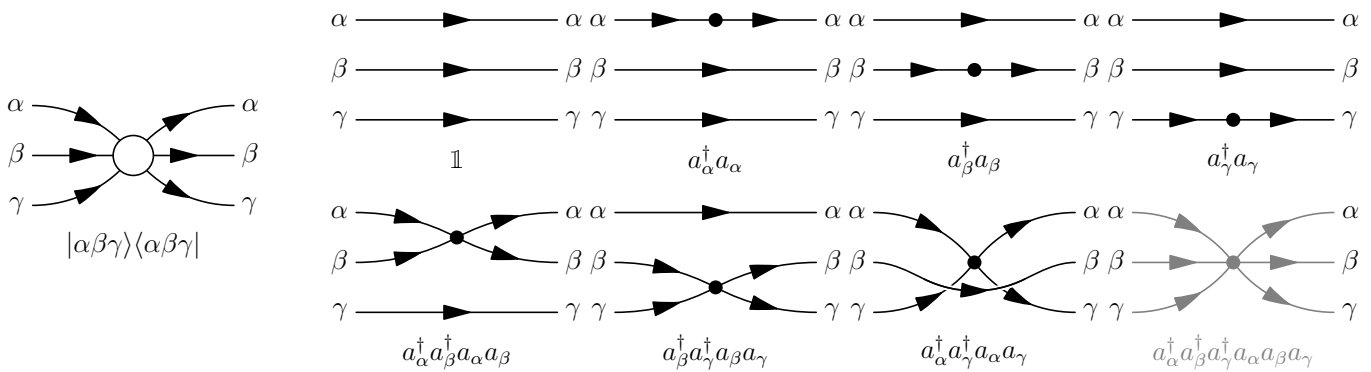}
\caption{\label{img:feynman}
  Decomposition of the diagonal element for the 
  three-particle state $\left|\alpha\beta\gamma\right\rangle$ into 
  irreducible interaction 
  processes. By truncation of three-particle processes, only the 
  three-particle irreducible term  $a^\dagger_\alpha a^\dagger_\beta 
  a^\dagger_\gamma a_\alpha a_\beta a_\gamma$ is neglected.}
\end{figure*}

One way to solve the flow equation \eqref{eq:flowequation} is to
parameterize the Hamiltonian in second quantization by an
adequate operator basis. 
In this way, the differential equations for the coefficients of the operators 
are found by computing the commutator on the right hand side 
in \eqref{eq:flowequation} and re-expressing the result
again in the operator basis chosen. The evaluation of the commutator generates
new many-body interaction processes that are not present in the initial 
Hamiltonian. They have to be incorporated in the Hamiltonian which results in 
an iterative calculation of the commutator.
This proliferation of terms generically yields an infinite number of 
differential equations which is intractible in practical applications. Four 
different strategies have evolved to  obtain a closed the set of
differential equations:

On restricting to finite systems, the flow equation can be solved exactly both 
on the level of second quantization or on the level of matrix elements. 
As a drawback, this 
approach is subjected to the same limitation as other finite-size methods. 
Furthermore, the differential equation system may still be large enough to 
require further approximations for practical applications.

In some special cases, it is possible to obtain a closed system of 
equations by identifying a small expansion parameter. In his original work 
\cite{Wegner1994}, Wegner investigated
the $n$-orbital model. He was able to close the system of differential 
equations in the limit of infinite $n$. However, these 
advantageous cases do not 
show up in every system or are simply out of interest. In general, 
approximations have to
be applied to the system to overcome the problem of proliferation of terms.

In the method of \emph{perturbative continuous unitary transformations} 
(P-CUT) introduced by Knetter and Uhrig 
\cite{uhrig98c,Knetter2000,Knetter2003},
the non-diagonality $H_\text{nd}$ is considered as small
perturbation to the diagonal part $H_\text{d}$  of the bare Hamiltonian. 
In this way, the flow equation can be expanded and therefore used to
apply perturbation theory up to very high orders.

The  non-perturbative approach, which we will use  in 
this work, is dubbed
\emph{self-similar continuous unitary transformations} 
(S-CUT).  Here a truncation scheme is defined
 that incorporates all terms considered to be important
to the problem while other terms are neglected. In view of the numerous 
degrees of freedom, the choice of an adequate truncation scheme is a 
non-trivial task.  It has to respect the system's physical properties. Since 
the structure of the Hamiltonian does not change
during the flow, it is named self-similar. As a rule of thumb, a
calculation is considered to be the more reliable the more terms are included 
in the truncation scheme.
Due to truncation, the modified flow equation reads
\begin{align}
 \partial_\ell H(\ell) = \widehat T \left[ \eta(\ell) , H(\ell) \right].
\label{eq:cut-truncflow}
\end{align}
Here we introduced the superoperator $\widehat T$ which denotes the 
application of the truncation scheme.

A natural description for many-body problems is provided by second 
quantization. Since most of the low-energy physics can be preserved by low 
numbers of suitable quasiparticles, it is useful to define a truncation scheme 
neglecting all terms that create or annihilate more than a given number of 
quasiparticles. We emphasise that this truncation expressed second 
quantization does not imply any restriction of the Hilbert space
which is to be considered a major advantage.

We stress that the action of a Hamiltonian on a state containing $n$ 
quasiparticles can be split into a sum of irreducible terms affecting 
at most $m\le n$ quasiparticles each (see Fig.\ \ref{img:feynman}).
Thus, the truncation of high-particle irreducible processes does not imply the 
complete neglect of matrix elements between states of high quasiparticle 
numbers, but rather an extrapolation based on lower quasiparticle 
irreducible processes.

In extended systems, often additional truncation criteria have to be applied
in order to close the set of differential equations. An obvious
choice for gapped systems with finite correlation lengt is to truncate 
according to the  real-space range of the physical process generated
by the term under study. We make use of 
this real-space truncation in Sect.\ \ref{struct:extend}.

Truncations cause quantitative errors in the calculated
physical quantities and may possibly lead to divergences of the truncated flow,
 even though convergence has been proven for the untruncated flow. 
In their analysis of the flow equations for the Anderson
model \cite{Kehrein1994}, Kehrein and Mielke used a strict truncation
scheme that includes only contributions of types that are already present 
in the bare Hamiltonian. 

In a second step, they included a new contribution to the scheme
and analyzed the new equations in order to assess its relevance. In their 
classification, terms that affect other matrix elements only by quantitative
deviations and vanish in the limit
$\ell\to\infty$ are denoted as irrelevant, or as marginal if they
converge to a finite value.
Only terms that are able to change the behavior of the flow equations
qualitatively, for instance that cause divergences if they are not treated 
properly, were considered to be relevant. 
This approach is suited to ensure the correct qualitative 
behavior of the effective model derived from the CUT. But
it does not provide a quantitative measure of the truncation errors.

\subsection{Symmetries\label{struct:cut-sym}}

For practical computations, exploiting symmetries of the Hamiltonian 
is generally very useful. To study Hamiltonians of systems with an infinite
 number of sites, the use of translation symmetry is inevitable.
Let us suppose that we have chosen a basis of operators to 
express the Hamiltonian. Each term in the Hamiltonian is given
by one of these basis operators multiplied by a prefactor, its coefficient.

Although the Hamiltonian is symmetric under a group of specific symmetry 
transformations, the symmetry of each individual basis operator may be lower. 
In  this case, the coefficients of several basis operators 
fulfil linear conditions to ensure that their combination is 
invariant under the symmetry group. By parametrizing the Hamiltonian 
in terms of linearly independent symmetric combinations of basis operators,
the number of coefficients to be tracked is significantly reduced 
saving both computation time and memory. 
Technically this can be done by selecting one operator as a unique 
representative  from which the complete symmetric combination can be obtained 
by taking the sum $\sum_G$ over a specific subgroup of the symmetry group of 
the Hamiltonian.

We emphasize that in general one has to clearly distinguish between 
the symmetry group of the Hamiltonian and the superoperator $\sum_G$.
Even if the Hamiltonian displays a continuous symmetry such as
the  $\text{SU}(n)$ spin rotation symmetry, the number of constraints 
that can be derived for the coefficients of the Hamiltonian's terms is
limited. Excluding translation 
symmetries, this means that only a finite number of terms
are represented by one representative operator. Therefore the number
of symmetry operations to build the symmetric combination from a single 
representative is often
finite although the exploited symmetry is continuous.
In summary, we take the  superoperation $\sum_G$ as a technical tool to benefit
from the Hamiltonian's underlying full symmetry group.

Moreover, the precise meaning of $\sum_G$ depends on the 
representative under study.
As an example, a representative that shares the whole symmetry of the 
Hamiltonian, e.g., unity, is already identical to its corresponding symmetric 
combination. To this operator, $\sum_G$ acts as identity. The other limit 
is a representative operator
that does not share any of the Hamiltonian's symmetries.
The superoperator $\sum_G$ applied to such a representative has to generate
the fully symmetric combination of basis operators.

Since the sums $\sum_G$ occur on both sides of Eq.\ \eqref{eq:flowequation}, 
the modified flow equation can be reduced to a representative expression 
requiring a sum for only one argument of the  commutator.
Correction factors have to be introduced since different representative basis 
operators appearing in the commutator give rise to different
numbers of basis operators in the  fully symmetric combination
which they represent. Details on the implementation of symmetries in S-CUT 
can be found  in Ref.\ \onlinecite{Reischl2006}.
\footnote{In Ref.\ \onlinecite{Reischl2006}, 
$\sum_G$ denotes a sum over the maximal symmetry group for 
all representatives. In contrast, the superoperator 
$\sum_G$ stands in our notation
only for the specific symmetry operations needed to
generate the fully symmetric combination represented
by the representative basis operator to which $\sum_G$ is applied. 
This leads to correction factors for 
the symmetrized flow equation instead of correction factors for coefficients 
of representatives as used by Reischl.}

\section{Mathematical analysis of truncation errors\label{struct:math}}

\subsection{Effects of truncation\label{struct:math-effects}}

On truncating the flow equation, all information about the truncated terms is 
lost. It is common practice to neglect terms which are considered to be 
unimportant and to justify this \emph{a posteriori}.
But even if these terms are not subject of the intended analysis of the 
effective Hamiltonian, their omission leads to quantitative deviations 
for the coefficients of \emph{all} terms in the Hamiltonian because
they are linked by the differential equations. 
We stress that the commutator of the generator and a truncated term may 
result in terms that comply with  the truncation scheme, i.e.,
that we want to compute quantitatively.
Thus the loss of information cannot be limited to certain terms only. 
Generically, truncation  introduces errors in \emph{all} coefficients 
of the Hamiltonian.

Because of truncation, the transformation of a Hamiltonian $H(\ell)$ described 
by the truncated flow equation \eqref{eq:cut-truncflow} does not need to be 
unitary anymore. Therefore, the spectrum of $H(\ell)$ will be distorted during 
the flow. Physical quantities calculated based on the effective Hamiltonian
are affected by finite inaccuracies.
In the following sections \ref{struct:math-split}-E, we present a formalism 
to bound these errors rigorously.

An additional physical consequence of truncation can be derived for real-space 
truncation schemes which neglect interactions beyond a certain range 
$d>d_\text{max}$ 
 (see Sect.\ \ref{struct:cut-truncscheme}). Due to the 
formulation in second quantization, the S-CUT method is capable to handle 
infinite systems. Nevertheless, the truncation by range affects correlations 
on larger length scales.
We observed that the coefficients of representatives in a truncated infinite 
system and a truncated periodic system with a certain size 
$l\geq L_\text{fin}$ share 
the same set of differential equations, if the algebra is local. Any 
possibility to observe that the system size is actually finite is masked by 
the truncation scheme if the system size is at least 
$L_\text{fin}=3 d_\text{max}+1$.

Therefore also the intensive physical properties of an infinite system 
determined using S-CUT with truncation range $d_\text{max}$ are 
\emph{identical} to those of a finite system with a certain size 
$l\geq L_\text{fin}$ and periodic 
boundary conditions. The quantity $L_\text{fin}$ can be understood as an 
effective size introduced by the real-space truncation scheme.
The mathematical derivation including a numerical verification is given in 
Appendix \ref{struct:app-effsize}.

\subsection{Splitting the flow equation\label{struct:math-split}}

To isolate the truncation error of S-CUT, we start from 
the full flow equation
\begin{align}
 \partial_\ell H(\ell) = [\eta^\prime(\ell),H(\ell)]
\label{eq:fullflow}
\end{align}
for the unitarily transformed Hamiltonian $H(\ell)$ with 
$H(\ell=0)=H^{(\text{b})}$. The
generator $\eta^\prime$ appears instead of $\eta$ because in practice the
generator is determined from the truncated Hamiltonian $H^{\prime}$
and not from $H$, see Eq.\ \eqref{eq:gendef}.

We decompose $H$ into two parts: the solution $H^{\prime}$  from a truncated 
flow equation and the difference $H^{\prime\prime}=H-H^{\prime}$ between the 
truncated and the non-truncated  calculation. Next, the flow equation 
\eqref{eq:fullflow} can be split into the system
\begin{subequations}
\label{eq:flows}
\begin{align}
\partial_\ell H^{\prime}(\ell)  &=& 
\widehat T &[\eta^\prime(\ell),H^{\prime}(\ell)] 
\label{eq:tflow}\\
\partial_\ell H^{\prime\prime}(\ell) &=& (\mathbbm{1}-\widehat T) 
&[\eta^\prime(\ell),H^{\prime}(\ell)]+ 
[\eta^\prime(\ell),H^{\prime\prime}(\ell)]
\label{eq:iflow}
\end{align}
\end{subequations}
of differential equations for $H^{\prime}$ and $H^{\prime\prime}$. 
As initial conditions,  we choose
\begin{subequations}
\label{eq:initial}
\begin{align}
H^{\prime}(0) \ =&\ H^{(\text{b})} 
\label{eq:tinitial}\\
H^{\prime\prime}(0)\ =&\ 0\label{eq:iinitial}.
\end{align}
\end{subequations}
Obviously, the sum of the equations \eqref{eq:flows} 
with the initial condition \eqref{eq:initial} reproduce 
the  flow equation \eqref{eq:fullflow} with its initial condition.

Up to now, the generator $\eta^\prime$ is not specified.
Using the generator scheme $\widehat\eta$, we define the generator
\begin{align}
  \label{eq:gendef}
 \eta^\prime(\ell) = \widehat\eta[H^{\prime}(\ell)]
\end{align}
as a function of the \emph{truncated} Hamiltonian.
Note that this choice does not violate the unitarity of the 
transformation  because the generator $\eta^\prime$ 
continues to be manifestly antihermitian.

Equation \eqref{eq:tflow} provides a \emph{closed} set of differential 
equations for the 
coefficients of the truncated Hamiltonian which can be treated by numerical 
integration. This leads to an effective Hamiltonian 
$H^{\prime}(\infty)=H^{\prime (\text{r})}$ with a structure 
determined by the chosen generator scheme.

In contrast, the full Hamiltonian $H$ is transformed by a true unitary 
transformation, but does not need to have any special structure in the limit 
of infinite $\ell$, since it is transformed like an observable by 
$\eta^\prime$. However, it is to be expected that it is close to 
$H^{\prime}$ if truncation errors are small.

The difference  $H^{\prime\prime}$ stores the complete 
\grq non-unitarity\grq\ of the transformation of $H^{\prime}(\ell)$. 
Mathematically,  Eq.\ \eqref{eq:iflow} describes a transformation of 
$H^{\prime\prime}$ via a flow equation with an additional \emph{inhomogeneity}
\begin{align}
 \kappa(\ell) = (\mathbbm{1}-\widehat T) 
&[\eta^\prime(\ell),H^{\prime}(\ell)]\label{eq:kappa}
\end{align}
depending on  $H^{\prime}(\ell)$. This natural emergence of an 
\emph{inhomogeneous flow equation} is quite remarkable and has not been
observed before to our knowledge. We emphasise that the number of equations 
defining $\kappa(\ell)$ remains finite if $H^{\prime}$ and thus $\eta^\prime$ 
are restricted by the truncation scheme to a finite number of terms.
Hence the computation of $\kappa(\ell)$ 
is indeed feasible. Of course, this is not true for $H^{\prime\prime}$.

\subsection{Inhomogeneous flow equation}

To solve the inhomogeneous flow equation \eqref{eq:iflow}, we use the ansatz
\begin{align}
 H^{\prime\prime}(\ell) = U(\ell)A(\ell)U^\dagger(\ell)
\end{align}
with $A(0)=H^{\prime\prime}(0)=0$. The unitary transformation $U(\ell)$ is 
linked to the generator $\eta^\prime(\ell)$ of the transformation by 
Eq.\ \eqref{eq:generator}. 
The formal solution for $U(\ell)$ using the $\ell$-ordering operator 
$\mathcal{L}$ reads
\begin{align}
U(\ell)=\mathcal{L}\exp \left(\int\limits_0^\ell \eta^\prime(\ell^\prime)
\mathsf{d} \ell^\prime \right).
\end{align}
Using variation of parameters
\begin{subequations}
\begin{align}
 \partial_\ell H^{\prime\prime}(\ell) &= 
\left[ \eta^\prime(\ell), H^{\prime\prime}(\ell)\right]+
U(\ell)\partial_\ell A(\ell)U^\dagger(\ell)
\\
&\overset{!}{=} \left[\eta^\prime(\ell),H^{\prime\prime}(\ell)\right]+
\kappa(\ell),
\end{align} 
\end{subequations}
leads to the equation
\begin{align}
 A(\ell) = A(0)+\int\limits_0^\ell U^\dagger(\ell^\prime)\kappa(\ell^\prime)
U(\ell^\prime)\mathsf{d} \ell^\prime.
\end{align}
Therefore, the formal solution of the inhomogeneous flow equation 
\eqref{eq:iflow} is given by
\begin{align}
 H^{\prime\prime}(\ell)= U(\ell)\left(H^{\prime\prime}(0)+\int\limits_0^\ell 
U^\dagger(\ell^\prime)\kappa(\ell^\prime)U(\ell^\prime)
\mathsf{d} \ell^\prime\right)U^\dagger(\ell).
\label{eq:isolved}
\end{align}
This expression \eqref{eq:isolved} has a very direct interpretation: 
All contributions 
of the inhomogeneity up to the given value of $\ell$ are re-transformed to 
$\ell=0$ and summed. This sum is evaluated after a unitary transformation to
the considered flow parameter.

\subsection{Truncation error}

The formal solution \eqref{eq:isolved} enables the calculation of the 
distortion of unitarity by the truncated calculation. All effects of the 
truncation are stored in $H^{\prime\prime}(\ell)$. Certainly, a direct 
calculation is neither practical 
nor desirable, because it is equivalent to an untruncated calculation. 
But for the derviation of a bound of the truncation error only a 
small part of the information is essential.
To assess the quality of the truncation, we are interested in the norm of 
$H^{\prime\prime}(\ell)$. In particular, we want to focus on norms 
that are unitarily invariant, i.e., 
\emph{invariant under unitary transformations}\footnote{It turned 
out that  the most appropriate choice is the spectral norm because it
implies  rigorous bound on eigen-values, see Sect.\ \ref{struct:math-bound}.} .

We apply the norm to Eq.\ \eqref{eq:isolved}. Due to its unitary invariance,  
we obtain
\begin{align}
  \left|\left|H^{\prime\prime}(\ell)\right|\right|=\left|
  \left|\int\limits_0^\ell 
  U^\dagger(\ell^\prime)\kappa(\ell^\prime)U(\ell^\prime)\mathsf{d} 
  \ell^\prime\right|\right|.
\end{align}
In addition, we used $H^{\prime\prime}(0)=0$ from Eq.\ \eqref{eq:iinitial}.
To avoid the complicated integration of an operator-valued function, 
we apply the triangle 
inequality to the Riemann integral arriving at the upper bound
\begin{subequations}
\begin{align}
\left|\left|H^{\prime\prime}(\ell)\right|\right|&\leq \int\limits_0^\ell 
\left|\left|U^\dagger(\ell^\prime)\kappa(\ell^\prime)U(\ell^\prime)\right|
\right|\mathsf{d} \ell^\prime
\\
&=\int\limits_0^\ell \left|\left|\kappa(\ell^\prime)\right|\right|\mathsf{d} 
\ell^\prime =: \Lambda_H(\ell),
\label{eq:def_Lambda_H}
\end{align}
\end{subequations}
where again the unitary invariance was used. We define 
the derived quantity $\Lambda_H(\ell)$ as \emph{truncation error} of the 
transformation. By construction, it is an upper bound for the distance between 
$H^{\prime}$ and $H$ measured by the selected norm. 
We emphasize that $\Lambda_H(\ell)$ is a scalar function which
depends only on the norm of the truncated terms as defined in Eq.\ 
\eqref{eq:kappa}. 
It starts at zero and increases monotonically with the flow parameter $\ell$.

Because of the finite and constant number of terms 
complying with the truncation scheme, the number of contributions to the 
inhomogeneity $\kappa(\ell)$ stays also constant
during the flow. The coefficients 
of the terms in $\kappa(\ell)$ can be calculated 
as functions of the coefficients of 
$H^{\prime}(\ell)$ already known by numerical integration.
This is the key simplification compared to the practically impossible 
direct  calculation of $H^{\prime\prime}(\ell)$ or $H(\ell)$.

In the above analysis, the necessary ingredients are the
flow equation and the truncation scheme. Hence
all considerations can also be carried over to the transformation of 
observables. The truncation error of an observable $\mathcal{O}$ can  be 
estimated analogously by
\begin{align}
  \Lambda_\mathcal{O}(\ell) := \int\limits_0^\ell \left|\left|(\mathbbm{1}-
  \widehat T) 
	   [\eta^\prime(\ell^\prime),\mathcal{O}^{\prime}(\ell^\prime)]
	   \right|\right|\mathsf{d} \ell^\prime \geq \left|
	   \left|\mathcal{O}^{\prime\prime}(\ell)\right|\right|
\end{align}
where $\mathcal{O}$ is decomposed in $\mathcal{O}^{\prime}$ and 
$\mathcal{O}^{\prime\prime}$ in analogy to \eqref{eq:initial}.
The generator is defined by the numerically accessible
truncated Hamiltonian $H^{\prime}$.

\subsection{Rigorous bounds for observables\label{struct:math-bound}}

The truncation error $\Lambda_\mathcal{O}$ is a property of the 
entire transformation of $\mathcal{O}^{\prime}$ that quantifies the 
loss of accuracy by truncation. It is desirable to have rigorous bounds for 
the accuracy of physical quantities calculated by truncated CUTs. 
Indeed, it is possible to obtain 
such bounds by calculating the truncation error defined by the spectral norm
\begin{align}
  \left|\left|A\right|\right|_S := \sqrt{\max\ \operatorname{EV}
    \left(A^\dagger A\right)}.
\end{align}
For hermitian operators, the spectral norm is identical to the maximum 
absolute eigen-value.

We denote the lowest eigen-value of $\mathcal{O}^{\prime}$ by 
$\Omega^\prime_\text{min}$ and 
the associate eigen-state by $\left|\psi^\prime\right\rangle$.
For the untruncated observable $\mathcal{O}$ we use $\Omega_\text{min}$ and 
$\left|\psi\right\rangle$. Since $\mathcal{O}$ is transformed by a unitary 
transformation, 
$\Omega_\text{min}$ does not change during the flow whereas 
$\Omega^\prime_\text{min}$ is changed due to truncation, for illustration 
see Fig.\ \ref{plot:toy-groundstate}.

The spectral norm of $\mathcal{O}^{\prime\prime}$ fullfills
\begin{subequations}
 \begin{align}
  \left|\left|\mathcal{O}^{\prime\prime}(\ell)\right|\right|  &\geq \left< \ 
  \mathcal{O}^{\prime\prime}(\ell) \ \right> {}_{{\psi^\prime(\ell)}}
  \\
  &= \underbrace{\left< \ \mathcal{O}(\ell) \ \right> 
    {}_{\psi^\prime(\ell)}}_{\geq\Omega_\text{min}} 
  - \underbrace{\left< \ \mathcal{O}^{\prime}(\ell) \ \right> {}_{
    }{\psi^\prime(\ell)}}_{\Omega_\text{min}^\prime(\ell)}.
 \end{align}
\end{subequations}
By condition, $\Omega_\text{min}$ is a lower bound for 
$\left< \ \mathcal{O}(\ell) \ \right> {}_{{\psi^\prime(\ell)}}$. It follows
\begin{align}
\left|\left|\mathcal{O}^{\prime\prime}(\ell)\right|\right| \geq  
\Omega_\text{min}- \Omega_\text{min}^\prime(\ell) =: 
\Delta \Omega_\text{min}(\ell).
\end{align}
Analogously, one obtains the inequality
\begin{subequations}
 \begin{align}
  \left|\left|\mathcal{O}^{\prime\prime}(\ell)\right|\right|  &\geq -\left< 
  \ \mathcal{O}^{\prime\prime}(\ell) \ \right> {}_{{\psi(\ell)}}
  \\
  &= -\underbrace{\left< \ \mathcal{O}(\ell) \ \right> 
    {}_{\psi(\ell)}}_{\Omega_\text{min}}  + 
  \underbrace{\left< \ \mathcal{O}^{\prime}(\ell) \ \right> 
    {}_{\psi(\ell)}}_{\geq
    \Omega_\text{min}^\prime(\ell)}.
 \end{align}
\end{subequations}
Since $\Omega_\text{min}^\prime(\ell)$ is a lower bound for 
$\left< \ \mathcal{O}^\prime(\ell) \ \right> {}_{{\psi(\ell)}}$, we obtain
\begin{align}
  \left|\left|\mathcal{O}^{\prime\prime}(\ell)\right|\right| \geq  -
  \Omega_\text{min}+ \Omega_\text{min}^\prime(\ell) 
  = -\Delta \Omega_\text{min}(\ell).
\end{align}

In summary, the truncation error
\begin{align}
  \Lambda_\mathcal{O}(\ell) \geq 
  \left|\left|\mathcal{O}^{\prime\prime}(\ell)\right|\right|\geq 
  \left|\Delta \Omega_\text{min}(\ell)\right|
\end{align}
is an upper bound of the deviation of the minimal eigen-value of the 
effective operator due to the truncation. Analogously, one can prove that 
$\Lambda_\mathcal{O}$ defines an upper bound for the deviation of the maximal 
eigen-value $\Delta\Omega_\text{max}$.

A very useful result ensues by considering the special case of the truncation 
error of the Hamiltonian itself
 because the ground state energy can directly be 
read off from the renormalized Hamiltonian $H^{(\text{r})}$.
Therefore, the exact ground state energy has to be within an interval of 
$\Lambda_H(\infty)$ around the ground state energy calculated by the 
truncated S-CUT
\begin{align}
\left|E_0-E_0^\prime\right|\leq \Lambda_H(\infty).
\end{align}
In this way, the truncation error defined by the spectral norm is no longer 
an abstract expression, but gives a practical error bound for a physical 
property of the system.

\section{Illustrative Model\label{struct:toy}}

\subsection{Double-Hard-Core-Boson}

As illustration of the formalism described above, 
we investigate the truncation error of a 
model of two sites which can be occupied by at most one particle each. To 
describe the system in second quantization, we use the hard-core boson 
language. The commutator of the associated annihilation and creation 
operators on site $i$ and $j$ is given by
\begin{align}
  \left[a_i,a^\dagger_j\right]=\delta_{ij} \left(\mathbbm{1} - a^\dagger_j 
  a_i\right) - a^\dagger_j a_i.
\end{align}

The Hamiltonian under study reads
\begin{subequations}
\label{eq:toy-dhard-core}
\begin{align}
H = 
\ \epsilon\mathbbm{1} &+ \mu \left(a_1^\dagger a_1 + a_2^\dagger a_2\right)+ 
t \left(a_1^\dagger a_2 + a_2^\dagger a_1\right)
\label{eq:toy-dhard-core-a}  \\
&+ \Gamma^{10}\left(a_1^\dagger + a_1 +a_2^\dagger +a_2\right)
\label{eq:toy-dhard-core-b} \\
& + \Gamma^{21}\left(a_1^\dagger a_2^\dagger a_2 
+a_1^\dagger a_1a_2^\dagger +h.c.\right)
\label{eq:toy-dhard-core-c}\\
& 
+ \Gamma^{20}\left(a_1^\dagger a_2^\dagger + a_1 a_2\right) + 
V a_1^\dagger a_1a_2^\dagger a_2.
\label{eq:toy-dhard-core-d}
\end{align}
\end{subequations}
Terms in the lines \eqref{eq:toy-dhard-core-c} and \eqref{eq:toy-dhard-core-d} 
are not present in the bare Hamiltonian but may emerge during the flow. 
The quantity $\epsilon$ defines the vacuum energy, $\mu$ stands for the 
chemical potential. The particle-particle interaction is denoted with $V$ and 
$t$ is the  prefactor of the hopping term. The quantities 
$\Gamma^{10}$, $\Gamma^{20}$ 
and $\Gamma^{21}$ violate the number of quasiparticles. They represent the 
non-diagonality of the Hamiltonian.

\subsection{Flow equations}

For our study, we use the particle conserving generator scheme 
$\eta^\prime(\ell)=\widehat \eta_{\text{pc}}[H^{\prime}(\ell)]$, 
for details see Refs.\ \onlinecite{uhrig98c,Knetter2000,Fischer2010}. The 
differential equations for the coefficients  of $H^{\prime}$ read
\begin{subequations}\begin{align}
    \partial_\ell \epsilon^ \prime  
    &=\! &\! -4&\Gamma^{10\prime}\Gamma^{10\prime}
    \! &\!-2&\Gamma^{20\prime}\Gamma^{20\prime}
    \\
    \partial_\ell \mu^\prime     	
    &=\! &\!  4 &\Gamma^{10\prime}\Gamma^{10\prime}
    \! &\!+2&\Gamma^{20\prime}\Gamma^{20\prime}
    \nonumber\\
    & \! &\! -2 &\Gamma^{21\prime}\Gamma^{21\prime}
    \! &\!-4&\Gamma^{10\prime}\Gamma^{21\prime}
    \\
    \partial_l t^\prime &=\! &\! -4 &\Gamma^{10\prime}\Gamma^{21\prime}
    \! &\!-2&\Gamma^{21\prime}\Gamma^{21\prime}
    \\
    \partial_\ell \Gamma^{10\prime} 
    &=\! &\!   -&\Gamma^{10\prime}\mu^\prime
    \! &\!- &\Gamma^{10\prime}t^\prime
    \nonumber\\
    & \! &\!   -&\Gamma^{20\prime}\Gamma^{10\prime}
    \! &\!-3&\Gamma^{20\prime}\Gamma^{21\prime}
    \\
    \partial_\ell \Gamma^{21\prime} &=\! &\!   -&\Gamma^{21\prime}\mu^\prime
    \! &\!+ &\Gamma^{21\prime}t^\prime&+2&\Gamma^{21\prime}
    \Gamma^{20\prime}
    \nonumber\\
    & \! &\!  +2&\Gamma^{10\prime}t^\prime \! &\!
    +4&\Gamma^{10\prime}\Gamma^{20\prime}	
    \\
    & \! &\!  -& \Gamma^{21\prime}V^\prime \! &\!
    -&	\Gamma^{10\prime}V^\prime
    \nonumber\\
    \partial_\ell \Gamma^{20\prime} 
    &=\! &\!  -2&\Gamma^{20\prime}\mu^\prime \! &\! -&
    \Gamma^{20\prime}V^\prime 
    \\
    \partial_\ell V^ \prime &=\! &\! 16& 
    \Gamma^{10\prime}\Gamma^{21\prime}\! &\! +8&
    \Gamma^{21\prime}\Gamma^{21\prime}.
\end{align}
\end{subequations}

Since the block-band structure is conserved by 
$\widehat\eta_{\text{pc}}$, see Ref.\
\onlinecite{Mielke1998,Knetter2000}, $\Gamma^{20\prime}$ 
stays zero during the flow unless it is already 
present in the initial Hamiltonian. As a mi\-ni\-mal truncation scheme, we 
neglect the particle-particle interaction given by 
$V^\prime$ in the following 
and thus the corresponding contributions to 
$\partial_\ell \Gamma^{21\prime}$ and $\partial_\ell \Gamma^{20\prime}$.
Therefore the only contribution to the inhomogeneity 
\begin{align}
  \kappa(\ell)= \left(\Gamma^{10\prime}(\ell)\Gamma^{21\prime}(\ell) +
  8\Gamma^{21\prime}(\ell)\Gamma^{21\prime}(\ell)\right) a_1^\dagger 
  a_1a_2^\dagger a_2
\end{align}
is given by the former derivative of the particle-particle interaction. 
To make use of the possibility of calculating a rigorous bound for the 
accuracy of the ground state energy and 
the maximal energy eigen-value, we choose the spectral norm. Since we use 
hard-core bosons, $a_1^\dagger a_1 a_2^\dagger a_2$ has the maximal 
eigen-value of unity. Equation \eqref{eq:def_Lambda_H} immediately yields the 
truncation error
\begin{align}
 \Lambda_H(\ell) = \int\limits_0^\ell 
\left|\Gamma^{10\prime}(\ell^\prime)\Gamma^{21\prime}(\ell^\prime) +
8\left(\Gamma^{21\prime}\right)^2(\ell^\prime)\right|\mathsf{d}\ell^\prime.
\end{align}

Due to the small number of couplings, we are able to calculate the quantities 
$H^{\prime}$ and $H^{\prime\prime}$ directly. 
To this end, we need to calculate $H(\ell)$ by 
transforming it like an observable 
under the flow of $\eta^\prime(\ell)$ following Eq.\ \eqref{eq:fullflow}. We 
stress that $H$ is transformed in this way by an exact unitary transformation 
without any truncations. We obtain the set of differential equations
\begin{subequations}
\begin{align}
  \partial_\ell \epsilon     	&=\! &\!  
  -4&\Gamma^{10\prime}\Gamma^{10}-2\Gamma^{20\prime}\Gamma^{20}
  \\
  \partial_\ell \mu     		&=\! &\!    
  &\Gamma^{10\prime}(4 \Gamma^{10}-2\Gamma^{21})+2\Gamma^{20\prime}
  \Gamma^{20}
  \nonumber\\
  & \! &\!  -2&\Gamma^{21\prime}(\Gamma^{10}+\Gamma^{21}) 
  \\
  \partial_\ell t     		&=\! &\!  
  -2&\Gamma^{10\prime}\Gamma^{21}-2\Gamma^{21\prime}(\Gamma^{10}+\Gamma^{21})
  \\
  \partial_\ell \Gamma^{10}   	&=\! &\!   
  -&\Gamma^{10\prime}(\mu+t+\Gamma^{20})-\Gamma^{21\prime}\Gamma^{20}
  \nonumber\\
  & \! &\!  -&\Gamma^{20\prime}(\Gamma^{10}+\Gamma^{21})
  \\
  \partial_\ell \Gamma^{21}   	&=\! &\!    
  &\Gamma^{10\prime}(2\Gamma^{20}+2t-V)+
  \Gamma^{20\prime}(2\Gamma^{10}+\Gamma^{21\prime})
  \nonumber\\
  & \! &\!   +&\Gamma^{21\prime}(-\mu+t-V+\Gamma^{20}) 
  \\
  \partial_\ell \Gamma^{20}   	&=\! &\!  
  -2&\Gamma^{10\prime}\Gamma^{21}+\Gamma^{20\prime}(-2\mu-V)+
  2\Gamma^{21\prime}\Gamma^{10} 
  \\
  \partial_\ell V     		&=\! &\!   
  8&\Gamma^{10\prime}\Gamma^{21}+8\Gamma^{21\prime}(\Gamma^{10}+\Gamma^{21}).
\end{align}
\end{subequations}
Thereby we are able to calculate $\left|\left|H^{\prime\prime}\right|\right|$ 
exactly as reference to estimate the quality of the truncation error 
$\Lambda_H$ which yields an upper bound to 
$\left|\left|H^{\prime\prime}\right|\right|$.
One should notice that the conservation of the bandstructure does not hold 
for $H$ since the generator depends on $H^{\prime}$.

\subsection{Results\label{struct:toy-res}}

\begin{figure}
 \includegraphics[width=\columnwidth]{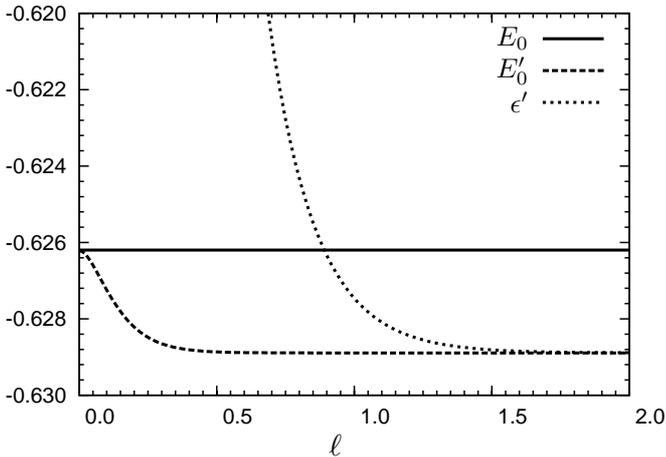}
 \caption{\label{plot:toy-groundstate}
The vacuum energy $\epsilon^\prime(\ell)$ of $H^{\prime}(\ell)$ 
converges to the ground state energy $E^\prime_0(\infty)$ of 
$H^{\prime}(\infty)$. Due to  truncation errors, the latter 
starts to deviate from the true ground state energy $E_0$ when the flow sets 
in at $\ell=0$. The calculation is carried out for 
$\mu^{(\text{b})}=2, t^{(\text{b})}=1$ and $\Gamma^{10(\text{b})}=1$.
}
\end{figure}

\begin{figure*}
\begin{minipage}{0.49\textwidth}\includegraphics[width=\textwidth]{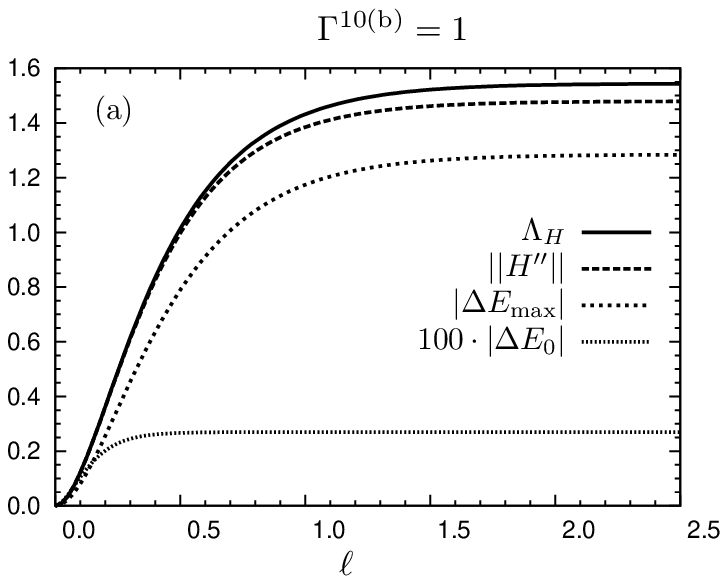}
\end{minipage}
\begin{minipage}{0.49\textwidth}\includegraphics[width=\textwidth]{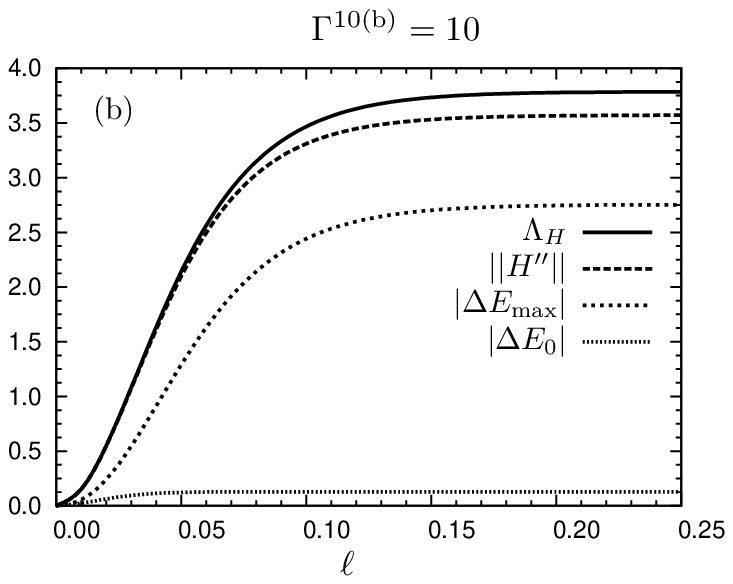}
\end{minipage}
\begin{minipage}{0.49\textwidth}\includegraphics[width=\textwidth]{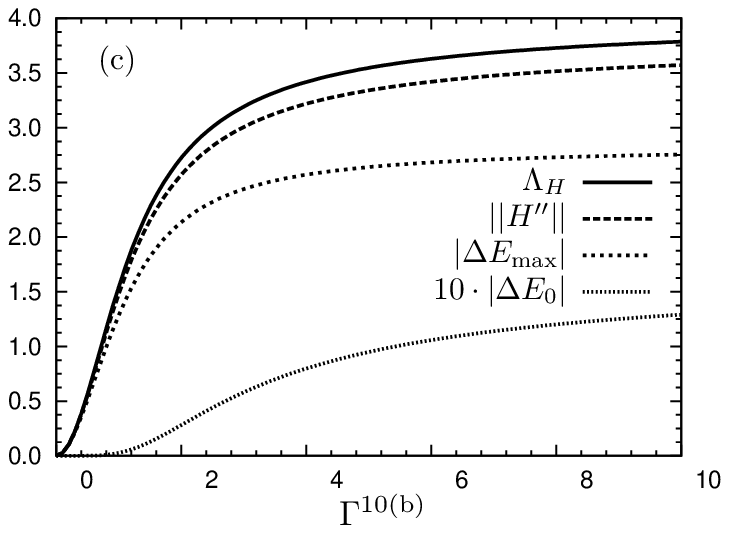}
\end{minipage} 
\begin{minipage}{0.49\textwidth}\flushleft
\caption{\label{plot:toy-errors}\hspace{23cm}} 
Truncation error $\Lambda_H$, spectral distance 
$\left|\left|H^{\prime\prime}\right|\right|$, deviation 
$\Delta E_\text{max}$ of the highest and lowest energy eigen-value 
$\Delta E_0$ of the truncated Hamiltonian for $\mu^{(\text{b})}=2$ and 
$t^{(\text{b})}=1$.
Panels a and b: dependence of the flow parameter $\ell$ for a moderate 
non-diagonality $\Gamma^{10{(\text{b})}}=1$ (a) and for large non-diagonality 
$\Gamma^{10{(\text{b})}}=10$ (b). 
Panel c: dependence of the renormalized quantities on the initial 
non-diagonality $\Gamma^{10{(\text{b})}}$.
\end{minipage}
\end{figure*}

For our calculations, we used $\mu^{(\text{b})}=2$ and $t^{(\text{b})}=1$ as 
initial conditions for $H^{(\text{b})}$, while $\epsilon^{(\text{b})}$, 
$\Gamma^{20{(\text{b})}}$ and 
$\Gamma^{21{(\text{b})}}$ are chosen to be zero. The initial non-diagonality 
$\Gamma^{10{(\text{b})}}$ is used to control the degree of necessary 
transformation.
For small values of $\Gamma^{10}$, the Hamiltonian is close to diagonality and 
therefore only slightly changed by the CUT. A large initial non-diagonality on 
the other hand requires intensive re-ordering processes in which truncation 
errors are important. 

Figure \ref{plot:toy-groundstate} shows the generic 
behavior of the ground state
 energy $E^{\prime}_0$ of ${H^{\prime}}^{(\text{r})}$ and the vacuum energy 
$\epsilon$ under the truncated flow. Truncation errors have a noticable impact 
due to the (large) non-diagonality $\Gamma^{10}$. 
In an early stage of the flow ($\ell \lesssim 0.5$), $E^{\prime}_0(\ell)$ 
starts to depart from $E_0$ and remains constant for the rest of the flow. 
The vacuum energy $\epsilon^\prime$ converges rapidly to the ground state 
energy $E^\prime_0(\infty)$ of $H^{\prime}(\infty)$.

The truncation error $\Lambda_H$ and the spectral distance 
$\left|\left|H^{\prime\prime}\right|\right|$ both 
saturate in the course of the truncated flow, see 
Fig.~\ref{plot:toy-errors}a and b, and 
they show a strong monotonic dependence of the non-diagonality. 
The spectral distance $\left|\left|H^{\prime\prime}\right|\right|$ is bounded 
by the truncation error as
it has to be. Furthermore, the truncation error turns out to be also a good 
approximation for the spectral norm. This can be seen in 
Fig.~\ref{plot:toy-errors} (panel c). Their difference is insignificant for 
small non-diagonalities and even for very large ones 
($\Gamma^{10{(\text{b})}}=10$) it takes only $6\%$.

The influence of truncation on the spectrum of $H^\prime$ is studied by the 
difference of the ground state energy $\Delta E_0=E_0-E^\prime_0$ and by the 
difference of the energy of the highest excited level $\Delta E_\text{max}=
E_\text{max}-E^\prime_\text{max}$ compared to the values of the initial 
Hamiltonian $H_0$. Both quantities stay clearly below the spectral distance, 
but differ in magnitude. For small values of 
$\Gamma^{10{(\text{b})}}$, $\Delta E_0$ is negligible and rises only up to 
$3.5\%$ of 
$\left|\left|H^{\prime\prime{(\text{r})}}\right|\right|$ for 
$\Gamma^{10{(\text{b})}}=10$. 
By contrast, $\Delta E_\text{max}$ is nearly identical to 
$\left|\left|H^{\prime\prime{(\text{r})}}\right|\right|$ for 
low non-diagonality and remains close to 
$\left|\left|H^{\prime\prime{(\text{r})}}\right|\right|$ 
even for high values of $\Gamma^{10{(\text{b})}}$. 
Thus the major impact of truncation occurs for the highest excited level.

This can be understood on recalling that the minimal truncation 
of the density-density interaction expressed by $V$ means neglecting the 
energy correction for the doubly occupied state and to approximate it by 
$E_0+2\mu$. Because this is precisly the highest eigen-state for 
vanishing non-diagonality $\Gamma^{10{(\text{b})}}$, it is directly affected 
by the truncation. In contrast, the ground state is only influenced indirectly 
by inaccuracies for the higher levels. Therefore, the low energy properties 
can be characterized by S-CUT very accurately despite of 
truncation, whereas larger inaccuracies occur at high energies.

In summary, the  truncation error $\Lambda_H$ defined in Eq.\ 
\eqref{eq:def_Lambda_H} is illustrated as an upper bound for the spectral 
distance $\left|\left|H^{\prime\prime}\right|\right|$ and for the errors of 
$E_0$ and $E_\text{max}$. We stress the good agreement of 
$\Lambda_H$, $\left|\left|H^{\prime\prime}\right|\right|$ and 
$\Delta E_\text{max}$. The error of ground state energy, however, is 
significantly lower than the bound given by 
$\left|\left|H^{\prime\prime}\right|\right|$. This can be explained as a 
particularity of the minimal truncation scheme which affects mainly high 
energies. We highlight that the truncation error measures the non-unitarity 
of the \emph{complete} transformation acting on the whole Hilbert space.

\section{Extended system\label{struct:extend}}

 \subsection{Dimerized spin-$\nicefrac{1}{2}$-chain\label{struct:ext-model}}

\begin{figure}
 \includegraphics[width=\columnwidth]{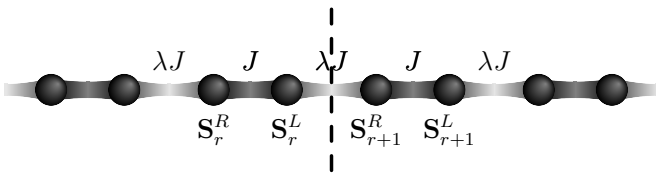}
 \caption{\label{img:chain}Schematic representation of the dimerized spin 
   chain. Dark bonds stand for the coupling $J$ between two 
   $S=\nicefrac{1}{2}$ spins forming a dimer, light bonds denote the variable 
   inter-dimer coupling $\lambda J$.
   The dashed line indicates an axis of reflection symmetry.}
\end{figure}

In the previous section, we studied the truncation error for a zero-dimensional
illustrative model. To illustrate the applicability of the 
truncation error for more relevant models, we extend our analysis to
a one-dimensional system. Furthermore, this allows us to calculate and to 
compare the truncation errors of more complex real-space truncation schemes.
As we will see, we have to use the triangle inequality again
to arrive at error bounds in extended systems.

The model studied is the one-dimensional dimerized antiferromagnetic spin 
$S=\nicefrac{1}{2}$ Heisenberg chain with the Hamiltonian
\begin{align}
  H=J\sum\limits_r \left(\textbf{S}_{r}^L\cdot\textbf{S}_{r}^R+
  \lambda\textbf{S}_{r}^R\cdot \textbf{S}_{r+1}^L\right), \quad J>0
  \label{eq:spinhammi}.
\end{align}
In this notation, $\textbf{S}_{r}^L$ stands for the operator of the left spin 
in the 
dimer on position $r$ and  $\textbf{S}_{r}^R$ for the right spin, 
see Fig.\ \ref{img:chain}. The parameter $0\leq\lambda\leq 1$ denotes the 
relative strength of interdimer coupling. It is used as a control parameter 
similar to $\Gamma^{10}$ in Sect.\ \ref{struct:toy}.

In the limit of $\lambda=0$, the system consists of isolated dimers. The 
ground state is given by a product state of singlets with a ground state 
energy of $\frac{E_0}{N}=-\frac{3}{4}J$ per dimer. The local $S=1$ 
excitations form equidistant spectrum with an increment of $J$.

For rising interdimer coupling, the excitations can be described by 
gapped spin $S=1$  quasiparticles called triplons \cite{Schmidt2003}. 
They can be seen as triplets with magnetic polarization cloud.
In the limit $\lambda=1$, the gap closes and the correlations decay 
algebraically \cite{cloiz66,yang66a,yang66b,fadde81,Klumper1998,Sachdev1999}. 
For $\lambda=1$, the system is the well-known homogeneous spin chain 
exactly solved 1931 by Bethe \cite{Bethe1931} by what was henceforth 
called Bethe ansatz. The ground state energy for the 
infinite chain was calculated by Hulth\'{e}n \cite{Hulthen1938} and takes the 
value $\nicefrac{1}{2}-2\ln 2$ per dimer and $J$.

With respect to the limit of isolated dimers, we choose for dimer $r$ a local 
basis of singlet/triplet states
\begin{subequations}
\begin{alignat}{3}
  \left|s\right\rangle_r &=  \frac{1}{\sqrt{2}} 
  \left(\left|\uparrow\downarrow\right\rangle-
  \left|\downarrow\uparrow\right\rangle\right)   	\	& &\ &      &
  \\
  \left|x\right\rangle_r &=  \frac{-1}{\sqrt{2}}
  \left(\left|\uparrow\uparrow\right\rangle  -
  \left|\downarrow\downarrow\right\rangle\right) 	\	&=&\ 
  t^\dagger_{x,r}&\left|s\right\rangle_r &
  \\
  \left|y\right\rangle_r &=  \frac{i}{\sqrt{2}} 
  \left(\left|\uparrow\uparrow\right\rangle  +
  \left|\downarrow\downarrow\right\rangle\right)	\	
  &=&\ t^\dagger_{y,r}&\left|s\right\rangle_r &
  \\
  \left|z\right\rangle_r &=  \frac{1}{\sqrt{2}} 
  \left(\left|\uparrow\downarrow\right\rangle+
  \left|\downarrow\uparrow\right\rangle\right)		\	&=&\ 
  t^\dagger_{z,r}&\left|s\right\rangle_r	&
\end{alignat}
\end{subequations}
as in the bond operator representation 
\cite{Chubukov1989-original,Chubukov1989,Sachdev1990}.
We define the reference state as the product state of singlets on each site
\begin{align}
 \left|0\right\rangle:=\bigotimes\limits_r \left|s\right\rangle_r.
\end{align}
The triplet operators are defined by
\begin{subequations}
  \begin{align}
    t^\dagger_{\alpha,r}&:=\left|\alpha\right\rangle_r\left\langle s\right|_r
    \\
    t^{\phantom{\dagger}}_{\alpha,r}&:=
    \left|s\right\rangle_r\left\langle\alpha\right|_r.
  \end{align}
\end{subequations}
Non-physical artifacts (e.g. states with two triplets on one dimer) are 
excluded. Therefore, the triplet operators obey the hard-core algebra
\begin{align}
  \left[t^{\phantom{\dagger}}_{\alpha,r},t^\dagger_{\beta,s}\right]=
  \delta_{rs}\delta_{\alpha\beta}\left(\mathbbm{1} - 
  \sum\limits_\gamma t^\dagger_{\gamma,r}
  t^{\phantom{\dagger}}_{\gamma,r}\right)-
  \delta_{rs}t^\dagger_{\beta,r}t^{\phantom{\dagger}}_{\alpha,r}.
\end{align}
The normal-ordered products of triplet operators (monomials) together with the 
identity $\mathbbm{1}$ form the basis $\{A_i\}$ for all operators on the 
lattice.  In this notation, the Hamiltonian reads
\begin{align}
  \label{eq:Whemmi}
  H = \sum\limits_r
  &\ -\frac{3}{4}\mathbbm{1} + t^\dagger_{\alpha,r}  
  t^{\phantom{\dagger}}_{\alpha,r}
  \nonumber\\
  &\ + \frac{1}{4}\lambda\left(t^\dagger_{\alpha,r} + 
  t^{\phantom{\dagger}}_{\alpha,r}\right)\left(t^\dagger_{\alpha,r+1} + 
  t^{\phantom{\dagger}}_{\alpha,r+1}\right)
  \nonumber\\
  &\ + \frac{i}{4}\lambda\epsilon_{\alpha\beta\gamma}
  \left(\left(t^\dagger_{\alpha,r} + 
  t^{\phantom{\dagger}}_{\alpha,r}\right)t^\dagger_{\beta,r+1}
  t^{\phantom{\dagger}}_{\gamma,r+1}\right. 
  \\
  &\ \left. - 	t^\dagger_{\beta,r}t^{\phantom{\dagger}}_{\gamma,r}
  \left(t^\dagger_{\alpha,r+1} + 
  t^{\phantom{\dagger}}_{\alpha,r+1}\right)\right)
  \nonumber\\
  + \frac{1}{4}\lambda&\left(t^\dagger_{\beta,r}
  t^{\phantom{\dagger}}_{\gamma,r}t^\dagger_{\gamma,r+1}
  t^{\phantom{\dagger}}_{\beta,r+1}-t^\dagger_{\beta,r}
  t^{\phantom{\dagger}}_{\gamma,r}t^\dagger_{\beta,r+1}
  t^{\phantom{\dagger}}_{\gamma,r+1}\right).&\nonumber
\end{align}
By the CUT the triplet states are mapped to re-normalized $S=1$ 
excitations (triplons). 

The Hamiltonian \eqref{eq:Whemmi} has three different 
symmetries that can be used for simplification of the
calculation as mentioned in Sect.\ \ref{struct:cut-sym}:

(i) The Hamiltonian \eqref{eq:Whemmi}  is self-adjoint. Although this is not a 
symmetry in the strict sense of the word, it implies  an additional constraint 
for the coefficients of $H$. Since quasiparticle creating and annihilating 
terms have to  occur in pairs in any Hamiltonian, one of them can be chosen as 
representative for the pair. Thereby the number of coefficients to be tracked 
is reduced.

(ii) The Hamiltonian \eqref{eq:Whemmi} shares the reflection symmetry of the chain 
$r\to -r$, see Fig.\ \ref{img:chain}. 
In addition, all left-spin and right-spin 
   operators have to be swapped implying 
 $t^{\phantom{\dagger}}_{\alpha,r}\leftrightarrow-t^{\phantom{\dagger}}_{\alpha,r}$ in the triplon notation
   \footnote{The choice of the reflection symmetry axis is arbitrary because 
     all reflection axes are equivalent due to translation symmetry.}.

(iii) The Hamiltonian \eqref{eq:spinhammi} is invariant under SU(2) rotations 
   in  spin space. Due to this invariance, the Hamiltonian written in the 
   triplet algebra \eqref{eq:Whemmi} can be decomposed
into symmetric combinations of terms. The terms in a symmetric combination 
differ only by permutations of triplet polarizations up to a sign factor.
The superoperator $\sum_G$  to build the symmetric combination of a maximally
asymmetric representative reads
\begin{align}
 \sum_{xyz}=\sum_{xy}\sum_{\text{cyc}} = \left(\mathbbm{1}+
\widehat S_{xy} \right)
 \left(\mathbbm{1}+\widehat S_\text{cyc}+\widehat S_\text{cyc}^2 \right),
\end{align}
where the cyclic permutations  of triplet polarizations is denoted
by $\widehat S_\text{cyc}$ and the exchange of the polarizations 
$x$ and $y$ with a negative 
sign factor for each triplet operator reads $\widehat S_{xy}$.

\subsection{Real-space truncation\label{struct:ext-trunc}}

For an extended system, the omission of processes that create or 
annihilate more than $N$ triplons as mentioned in Sect.\
\ref{struct:cut-truncscheme} can be insufficient because the number of 
remaining terms is still infinite. 
For example, even by restricting to processes of at most two 
triplons, an  infinite number of independent terms varying by range can emerge
in the course of the flow. Since an energy gap 
implies a finite correlation length, 
it is an adequate choice to neglect all processes that exceed a given range 
$d_\text{max}$. 
In a one dimensional system, the range can easily  be defined as the 
distance of the rightmost and leftmost triplet operator in a term.

In particular, it has turned out to be advantageous to use a combination 
of both the quasiparticle and the range criterion as truncation scheme 
\cite{Reischl2006,Fischer2010}.
For the most important  processes of low quasiparticle number, e.g., the
 hopping of triplons, a long range is allowed for to preserve most of the 
relevant physics. For more complex processes of more quasiparticles
only a shorter range can be considered 
because their number increases much more  steeply with range. 
But since more quasiparticles are required for such terms to
become active the reduced range does not need to imply a reduced
accuracy.

In view of the above considerations
 we classify terms by the sum $n$ of created and annihilated 
quasiparticles. For each value of $n$ a specific maximal range $d_n$ is 
defined. The complete truncation scheme can be written as 
$\textbf{d}=\left(d_2,d_3 \dots d_{2N}\right)$ where at most $N$ quasiparticles
 may be created or annihilated. No maximal range needs to 
be specified for $n=1$ because those terms always have 
range zero and their number 
is, due to translation symmetry, restricted to the six local creation 
and annihilation operators. Furthermore, without magnetic field
no single annihilation or creation of triplons takes place
due to the conservation of the total spin.

\subsection{Triangle inequality\label{struct:ext-tri}}

To calculate the truncation error according to Eq.\ \eqref{eq:def_Lambda_H}, 
the norm of the inhomogeneity
\begin{align}
\kappa(\ell)=
\sum\limits_{A_i}\kappa_{A_i}(\ell)A_i
\end{align}
 has to be calculated.
Its coefficients $\kappa_{A_i}$ are obtained using Eq.\ \eqref{eq:kappa} from 
the coefficients of $H^{\prime}$ by evaluating the terms of the commutator 
that are discarded in the truncation scheme.
The precise calculation of $\left|\left|\kappa\right|\right|$ is not feasible 
for large systems because the effort to calculate the maximal eigen-value is 
too large. Since we are only interested in an upper bound, we apply the 
triangle inequality again to reach
\begin{align}
  \left|\left|\kappa(\ell)\right|\right|=\left|\left|
  \sum\limits_{A_i}\kappa_{A_i}(\ell){A_i}\right|\right|\leq 
  \sum\limits_{{A_i}}\left|\kappa_{A_i}(\ell)\right|\left|\left|A_i\right|
  \right|.
\end{align}
In this way, we define an upper bound $\widetilde \Lambda_H$ 
or the truncation  error
\begin{align}
  \widetilde\Lambda_H(\ell)=\sum\limits_{{A_i}}\int\limits_0^\ell 
  \left|\kappa_{A_i}(\ell^\prime)\right|\mathsf{d} \ell^\prime
  \left|\left|A_i\right|\right|\geq \Lambda_H(\ell).
\end{align}
Therefore, $\widetilde\Lambda_H$ is also an upper bound for $\Delta E_0$ and 
$\Delta E_\text{max}$, although it is less strict than $\Lambda_H$.

Recall that our basis operators $A_i$ are normal-orderd products 
of hard-core-boson creation and annihilation operators. It turns out that in 
this case the spectral norm $\left|\left|A_i\right|\right|$ can be calculated 
easily since the product ${A_i}^\dagger {A_i}$ is already diagonal with respect
to the dimer eigen-states. It is a product of local triplon density terms.
Thus ${A_i}^\dagger {A_i}$ has the eigen-value unity for all configurations 
having triplons with the polarization of the local triplon density operators 
on each site of the cluster of ${A_i}$
\footnote{The cluster of a term is the set of sites on which its 
action differs from identity.}. 
For different occupations, ${A_i}^\dagger {A_i}$ has the eigen-value zero.
Therefore the spectral norm for all elements of the operator basis 
$\{A_i\}$ is unity.

\subsection{Optimization using symmetries\label{struct:ext-sym}}

As pointed out in Sect.\ \ref{struct:cut-sym}, exploiting symmetries 
reduces the computational effort for the S-CUT method significantly. 
In addition, symmetries can be used to reduce the bound 
$\widetilde \Lambda$. To see this, we write $\Lambda$ using the basis of 
representatives $\{C_i\}$ of operator monomials
\begin{subequations}
  \begin{align}
    \Lambda_H(\ell)
    &=    \sum\limits_{C_i} \int\limits_0^\ell 
    \left|\left|\kappa_{C_i}(\ell^\prime)\sum_G {C_i}\right|\right|\mathsf{d} 
    \ell^\prime
    \\
    &\leq \sum\limits_{C_i} \int\limits_0^\ell 
    \left|\kappa_{C_i}(\ell^\prime)\right|
    \mathsf{d} \ell^\prime \cdot \left|\left|\sum_G {C_i}\right|\right| =: 
    \widetilde\Lambda_H(\ell).
  \end{align}
\end{subequations}
We use the triangle inequality again to decompose 
$\left|\left|\sum_G {C_i}\right|\right|$ into the  weight $w_{C_i}$ which 
denote the number of terms generated by $\sum_G$  
multiplied by  the norm of the representative. This yields
\begin{align}
  \widetilde \Lambda_H(\ell) = \left(\int\limits_0^\ell 
  \left|\kappa_{C_i}(\ell^\prime)\right|\mathsf{d}\ell^\prime\right) 
  w_{C_i}\left|\left|{C_i}\right|\right|.
  \label{eq:def_Lambda_T_H}
\end{align}

If, however, we avoid the triangle inequality for the fully
symmetric combination of basis operators we obtain a better,
stricter upper bound. This enters Eq.\ \eqref{eq:def_Lambda_T_H} 
by replacing $w_{C_i}$ by a reduced effective weight $w^G_{C_i}$.
In order to determine this reduction, we have to calculate the relation 
between the norm of the fully symmetric combination and the norm of a 
single representative analytically
\begin{align}
  \left|\left|\sum_G {C_i}\right|\right|=:w^G_{C_i} 
  \left|\left|{C_i}\right|\right|.
\end{align}
The effective weight factor $w^G_{C_i}$ 
allows us to modify Eq.\ \eqref{eq:def_Lambda_T_H} to calculate the 
improved truncation bound  $\widetilde\Lambda^G$. 
In presence of multiple symmetries, it is possible to
improve the weight with respect to only some selected symmetries. The other 
symmetries can be treated by using the triangle inequality as done in Eq.\ 
\eqref{eq:def_Lambda_T_H}. They continue to enter the combined weight by an
additional  factor.

As an example, we consider the  self-adjointness to reach
the improved truncation bound $\widetilde\Lambda^\dagger_H$.
To determine its weight, it is useful to decompose the action of a 
representative
\begin{align}
  C=\left|c_1\right\rangle\left\langle c_2\right|
  \bigotimes\limits_{r \notin \text{cluster}}\mathbbm{1}
\end{align}
into the non-trivial action on its cluster and into 
 the identity on the rest of the system.
 Since $C$ is a normal-ordered product of hard-core-boson operators, 
the action on its cluster is given by only one non-vanishing matrix element.

To build the corresponding symmetric combination $\sum_G C$, we have to 
distinguish two cases:

(i) \emph{$\left|c_1\right\rangle=\left|c_2\right\rangle$}\quad
If $C$ is  self-adjoint, it is already a symmetric combination 
and thushas a weight factor of unity in both cases,
i.e., using and not using the self-adjointness.

(ii) \emph{$\left|c_1\right\rangle\neq\left|c_2\right\rangle$}\quad
In this case, the symmetric combination for $C$ reads
\begin{align}
  \sum_G C = \left(\mathbbm{1} + \widehat A\right) C = 
  \left(\left|c_1\right\rangle\left\langle c_2\right|+\left|c_2\right\rangle
  \left\langle c_1\right|\right) \bigotimes\limits_{r \notin \text{cluster}}
  \mathbbm{1}.
\end{align}
Due to our choice of basis, both $\left|c_1\right\rangle$ and 
$\left|c_2\right\rangle$ are eigen-states 
of local triplet density operators and therefore orthogonal, which implies
the maximal eigen-value of 1 for $\left|c_1\right\rangle
\left\langle c_2\right|+\left|c_2\right\rangle\left\langle c_1\right|$.
Hence  $\Vert\left|c_1\right\rangle\left\langle c_2\right|\Vert=1$ and 
$\Vert\left|c_1\right\rangle\left\langle c_2\right|+
\left|c_2\right\rangle\left\langle c_1\right|\Vert=1$ which implies
a gain of a factor of 2 if the triangle inequality is avoided which
would have led to $\Vert\left|c_1\right\rangle\left\langle c_2\right|+
\left|c_2\right\rangle\left\langle c_1\right|\Vert
\le \Vert\left|c_1\right\rangle\left\langle c_2\right|\Vert + 
\Vert\left|c_2\right\rangle\left\langle c_1\right|\Vert =2$.
As an example, we consider the representative 
$t^\dagger_{x,1}t^{\phantom{\dagger}}_{x,1}t^\dagger_{y,2}$. 
On its cluster $\{1,2\}$ it 
stands for a transition from $\left|x_1s_2\right\rangle$ to 
$\left|x_1y_2\right\rangle$.
The matrix representation for the action of its symmetric combination
\begin{align}
  \sum_G t^\dagger_{x,1}t^{\phantom{\dagger}}_{x,1}t^\dagger_{y,2} = 
  \left(\left|x_1y_2\right\rangle\left\langle x_1s_2\right|+
  \left|x_1s_2\right\rangle\left\langle x_1y_2\right|\right) 
  \bigotimes\limits_{r \notin \{1,2\}}\mathbbm{1}
\end{align}
has zero matrix elements except for a $2\times2$ block 
with the eigen-values -1 and +1.

In conclusion, using self-adjointness  to gain an effective weight 
$w^\dagger$ saves a factor of 2 for non-symmetric terms.

This effective weight can be improved further 
by considering spin symmetry. The complete symmetry group 
with respect to all permutations of triplet polarizations 
can be decomposed into the subgroup of cyclic permutations and the subgroup 
of the transposition of  $x$ and $y$ triplets $\widehat S_{xy}$ plus the 
identity. For simplicity, we concentrate on the latter one only and calculate 
the truncation bound $\widetilde\Lambda_H^{\dagger,xy}$.
The matrix associated with the $\sum_G C$ action on its cluster can have up to 
four non-vanishing matrix elements , i.e., four states of the cluster 
have to be  taken into account.
The representatives can be classified by the specific action
of $\sum_G$  needed to obtain the corresponding fully symmetric combination:

\emph{$\sum_G=\mathbbm{1}$:} These highly symmetric representatives 
(e.g.\ $t^\dagger_{z,1}t^{\phantom{\dagger}}_{z,1}$) are invariant under both 
$\widehat A$ and $\widehat S_{xy}$. They have weight $w^{\dagger,xy}$ unity.

\emph{$\sum_G=\mathbbm{1}+\widehat A$:} 
These representatives are not self-adjoint, but 
invariant under either transposition of $x$ and $y$ (e.g.\ 
$t^{\phantom{\dagger}}_{z,1}$) or 
under the combination of transposing and adjunction 
(e.g.\ $t^\dagger_{x,1}t^{\phantom{\dagger}}_{y,1}$). As discussed previously, 
the weight $w^{\dagger,xy}$ takes the  value of 1 instead of 2.

 \emph{$\sum_G=\mathbbm{1}+\widehat S_{xy}$:} 
For representatives that are self-adjoint, but not 
invariant under transposition (e.g.\ 
$t^\dagger_{x,1}t^{\phantom{\dagger}}_{x,1}$), the weight 
$w^{\dagger,xy}$ is reduced to the value 1 instead of 2 as well.

\emph{$\sum_G=\left(\mathbbm{1}+\widehat A\right)\left(\mathbbm{1}+
\widehat S_{xy}\right)$:}
In this asymmetric case, the norm of the symmetric combination can be 
either one (e.g. $t^\dagger_{x,1}t^{\phantom{\dagger}}_{y,2}$) or $\sqrt{2}$ 
(e.g.\ $t^\dagger_{x,1}$). Therefore,  $w^{\dagger,xy}=\sqrt{2}$ can be used
instead of $w=4$ .

The weights can be improved by exploiting further symmetries, e.g., 
cyclic spin permutations or reflection symmetry. But the complexity of the 
necessary case-by-case analysis rises considerably.
Especially the use of point group symmetries of the lattice is complicated
because the cluster of monomials linked by a point group do not need  to be 
identical. Thus the inclusion of point group 
symmetries in the calculation of the 
bounds is beyond the scope of this article.

 \subsection{Results for an infinite system\label{struct:inf-res}}

\begin{figure}
  \includegraphics[width=\columnwidth]{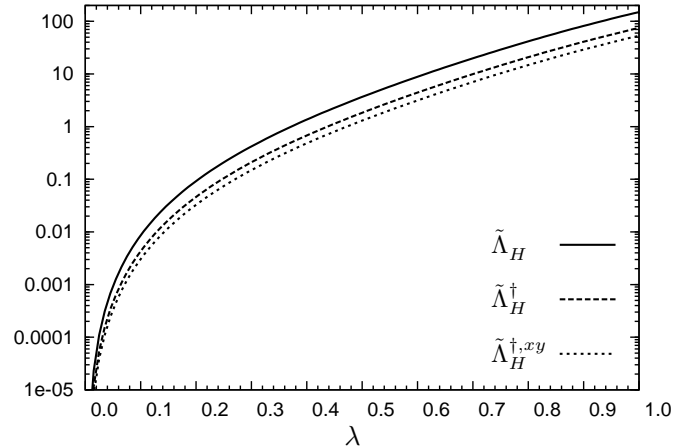}
  \caption{\label{plot:ext-863detail} 
    Renormalized truncation bound per dimer $\widetilde\Lambda_H$ and reduced 
    bounds $\widetilde\Lambda_H^\dagger$ (exploiting self-adjointness) and 
    $\widetilde\Lambda_H^{\dagger,xy}$ (exploiting self-adjointness and $xy$ 
    symmetry) vs.\ the interdimer coupling $\lambda$ for the ground state 
    generator using the truncation scheme $\textbf{d}=\left(8,6,6,3,3\right)$. 
  }
\end{figure}
\begin{figure}
  \includegraphics[width=\columnwidth]{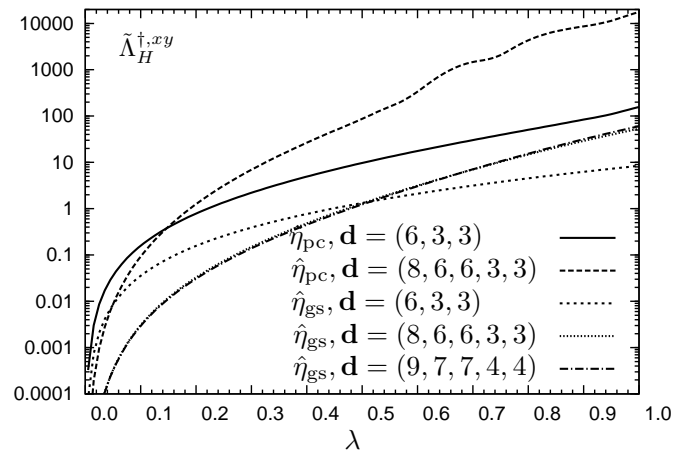}
  \caption{\label{plot:ext-infinite} 
    Renormalized reduced truncation bounds per dimer 
    $\widetilde\Lambda_H^{\dagger,xy}$ vs.\ interdimer coupling 
    $\lambda$ for ground state and particle conserving generator using 
    various truncation schemes. }
\end{figure}

\begin{table}
  \begin{tabular}{cl rrrr }
    \hline
    \multicolumn{1}{c}{$\widehat\eta$}	&	
    \multicolumn{1}{c}{$\textbf{d}$}	&	
    \multicolumn{1}{c}{$\left|\Delta E_0\right| $}	&	
    \multicolumn{1}{c}{$\widetilde\Lambda_H$}	&	
    \multicolumn{1}{c}{$\widetilde\Lambda_H^\dagger$}	&	
    \multicolumn{1}{c}{$\widetilde\Lambda_H^{\dagger, xy}$}	
    \\
    \hline
    $\widehat\eta_\text{pc}$	&$	(6,3,3)	$&	0.01739	&	
    439.37	&	219.93	&	156.74	
    \\
    $\widehat\eta_\text{pc}$	&$	(8,6,6,3,3)	$&	
    0.00032	&	49926.1\phantom{0}	&	24974.8\phantom{0}
    &	17660.2\phantom{0}	
    \\
    $\widehat\eta_\text{gs}$	&$	(6,3,3)	$&	0.02675	&	
    23.51	&	11.75	&	8.41	
    \\
    $\widehat\eta_\text{gs}$	&$	(8,6,6,3,3)	$&	
    0.00915	&	149.31	&	74.68	&	52.98	
    \\
    $\widehat\eta_\text{gs}$	&$	(9,7,7,4,4)	$&	
    0.00948	&	169.32	&	84.67	&	60.04	
    \\
    \hline
  \end{tabular}
  \caption{\label{tab:ext-infinite} Numerical values for the renormalized 
    truncation bound per dimer $\widetilde\Lambda_H$ and reduced bounds 
    $\widetilde\Lambda_H^\dagger$, $\widetilde\Lambda_H^{\dagger,xy}$ for 
    $\lambda=1$ for the generators and the truncation schemes used in Fig.\ 
    \ref{plot:ext-infinite}. The inaccuracies of the ground state energy 
    $\left|\Delta E_0\right|$ are calculated with respect to the analytical 
    result \cite{Hulthen1938}.}
\end{table}

Figure \ref{plot:ext-863detail} shows the bound $\widetilde\Lambda_H$ 
defined in Eq.\ \eqref{eq:def_Lambda_T_H} for the  infinite 
dimerized Heisenberg chain using the ground state generator scheme 
$\widehat\eta_\text{gs}$ \cite{Fischer2010} defined in \eqref{eq:cut-gs} and 
the truncation 
scheme $\textbf{d}=(8,6,6,3,3)$ as function of the interdimer coupling 
$\lambda$. For rising values of $\lambda$, the truncation bound increases 
drastically similar to the behavior observed for the illustrative model. 
But it attains a significantly higher absolute value finally.

For weak interdimer coupling $\lambda \lesssim 0.1$, the error bound for the 
ground state energy given by $\widetilde\Lambda$ 
is useful as a rigorous bound. 
But this estimate becomes inappropriate for medium and strong coupling since 
the $\widetilde\Lambda$ grows rapidly to the same magnitude as $E_0$ and 
beyond so that it does no longer represent a
meaningful bound.

For $\lambda=1$ the analytical result can be used as reference to determine 
the error of the ground state energy per dimer $\left|\Delta E_0\right|$ 
determined as the renormalized vacuum energy of the CUT, see 
Tab.\ \ref{tab:ext-infinite}. This relative error is only $1.04\%$. 
In contrast, the truncation bound  $\widetilde\Lambda_H$ 
exceeds the ground state energy by several orders of magnitude.

This discrepancy stems from the fact that the truncation error $\Lambda_H$ 
measures truncation effects of the entire transformation, 
not only of the inaccuracies of ground state energy in particular. 
The extensive use of the triangle inequality to
calculate the truncation bound $\widetilde\Lambda$ enhances this difference 
additionally, although it can be reduced exploiting symmetry.
The interesting question which effect dominates
is postponed to the study of a finite system below
where all quantities are numerically accessible.

The use of the adjunction symmetry reduces the bound by a factor of about two, 
and  the use of $xy$ symmetry reduces it by an additional factor of about 
$\sqrt{2}$. In both cases, exploiting the symmetry pays in decreasing the 
bound close to the optimum which can be achieved for terms with the lowest 
symmetry. Since terms with low symmetry occur much more frequently than 
symmetric ones, the error bound is reduced efficiently by exploiting 
symmetries. However, the gain achieved in this way
can not overcome the tremendous factor (orders of magnitude) between the 
upper bound and to the error of the ground state energy.

In the extended system, various truncation schemes of different quality and 
computational effort can be used. Figure \ref{plot:ext-infinite} shows the 
truncation bound $\widetilde\Lambda_H^{\dagger,xy}$ for different truncations 
and generator schemes. Besides the particle conserving generator
$\widehat\eta_\text{pc}$ the ground state decoupling generator 
$\widehat\eta_\text{gs}$ is applied, for details see Ref.\ 
\onlinecite{Fischer2010}.
The numerical values for $\lambda=1$ including the 
difference to the exact ground state energy are given in Tab.\ 
\ref{tab:ext-infinite}. 

In general, the particle conserving generator scheme $\widehat\eta_\text{pc}$
yields significantly higher truncation bounds than $\widehat\eta_\text{gs}$, 
although the inaccuracies of the calculated ground state energies are lower 
(using equal truncation schemes $\textbf{d}$).
This is due to the fact that $\widehat\eta_\text{pc}$  performs a more 
comprehensive reordering of the quasiparticle subspaces and includes much 
more terms than  $\widehat\eta_\text{gs}$. This implies a higher impact of 
truncation errors.
Much more terms emerge in the  evaluation of the commutator and have to be 
incorporated in $\kappa$ resulting in a larger differential equation system 
with much more coefficients.

The dependence on the truncation scheme is more complex. For weak coupling, 
looser truncation schemes imply lower truncation error bounds than stricter 
schemes. This is what one expects naively since the inclusion of more
and more terms, hence a less strict truncation, should describe
the system better and better. Thus the truncation error should decrease.

But this relation can be inverted for strong coupling
 \emph{although} the looser schemes 
reproduce the analytical result for the ground state energy with much higher 
accuracy for both generator schemes as seen for $(6,3,3)$ and $(8,6,6,3,3)$. 
We call this phenomenon the \emph{truncation paradoxon} because
it seems to be paradoxical at first glance. We stress that
it does not represent a logical contradiction but only a counterintuitive
behaviour. 

On second thought, one realizes that
the use of a looser truncation scheme increases the number of terms in 
$\kappa$ drastically. For instance, an increase from 30.972 for $(6,3,3)$ to 
16.777.215 representatives for $(8,6,6,3,3)$ is found in the 
$\widehat\eta_\text{pc}$
scheme. Irrespective of the consequences to the spectral norm of $\kappa$, 
this massive increase of terms has a big impact on the bound 
$\widetilde\Lambda$ which relies on the triangle inequality
to bound each of these terms separately. Hence the bound
is so large simply because it is very loose.

\begin{figure}
  \includegraphics[width=\columnwidth]{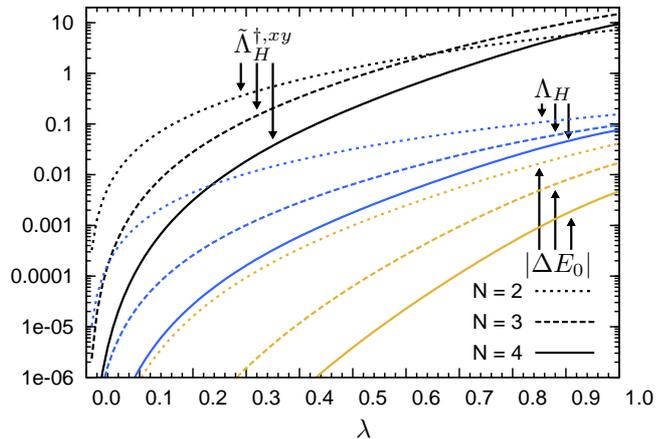}
  \caption{\label{plot:ext-finite}  (Color online)
    Renormalized truncation bound per dimer 
    $\widetilde\Lambda_H^{\dagger,xy}$, 
    exact renormalized truncation error per dimer $\Lambda_H$ and error of 
    ground state energy per dimer $\left|\Delta E_0\right|$ vs.\ interdimer 
    coupling $\lambda$ for the ground state generator considering at most 
    $N$ triplon operators per term.}
\end{figure}

\begin{table}
  \begin{tabular}{cl rrrr }
    \hline
    $N$	&	$\left|\Delta E_0\right| $	&	$\Lambda_H$
    &	$\widetilde\Lambda_H$	&	$\widetilde\Lambda_H^\dagger$
    &	$\widetilde\Lambda_H^{\dagger, xy}$	
    \\
    \hline
    2	&	0.04109	&	0.1556	&	20.34	&	10.17	&
    7.28	
    \\
    3	&	0.01709	&	0.0964	&	41.87	&	20.93	&
    14.86	
    \\
    4	&	0.00465	&	0.0768	&	26.78	&	13.39	&
    9.48	
    \\
    \hline
  \end{tabular}
  \caption{\label{tab:ext-finite} 
    Error of ground state energy per dimer $\left|\Delta E_0\right|$, exact 
    renormalized truncation error per dimer $\Lambda_H$ and renormalized 
    truncation bounds per dimer $\widetilde\Lambda_H^{}$, 
    $\widetilde\Lambda_H^{\dagger}$ and $\widetilde\Lambda_H^{\dagger,xy}$ for 
    $\lambda=1$ using the ground state generator scheme 
    $\widehat\eta_\text{gs}$ 
    considering at most $N$ triplon operators per term.}
\end{table}

 \subsection{Results for a finite extended system\label{struct:fin-res}}

The question arises whether the truncation paradox is caused by the extensive 
use of triangle inequality to determine the bound $\widetilde\Lambda$, 
or whether it is an intrinsic characteristics 
of the truncation error $\Lambda$ itself. To  investigate this issue, 
it is necessary to determine the exact truncation error 
$\Lambda$ without use of the triangle inequality. Thus an exact 
diagonalization of $\kappa$ is required.
This restricts us to the investigation of a finite chain segment.

In the following, we study the periodic, dimerized Heisenberg chain 
consisting of five dimers.
In view of the small size of the system, we use the maximal number $N$ of 
interacting triplons as only truncation criterion. No extensions are 
considered.

Figure \ref{plot:ext-finite} shows  the truncation bound 
$\widetilde\Lambda_H^{\dagger,xy}$, the exact truncation error $\Lambda$ and 
the deviation of the ground state energy $\left|\Delta E_0\right|$ vs.\ 
$\lambda$. The values  for $\lambda=1$ are given in Tab.\ \ref{tab:ext-finite}.
It turns out that \emph{both} the calculation of the exact truncation error 
$\Lambda_H$ and the extensive use of the triangle inequality contribute to the 
very large values of $\widetilde \Lambda_H^{\dagger,xy}$. For small values of 
$\lambda$, the large value of $\Lambda_H$ dominates the truncation bound 
$\widetilde\Lambda_H^{\dagger,xy}$ while for large values of $\lambda$, the 
approximation using the triangle inequality to bound the very many
arising terms contributes most.

In a direct comparision, the exact truncation error $\Lambda_H$ is 
overestimated by the bound $\widetilde\Lambda_H^{\dagger,xy}$ using the 
triangle inequality by two orders of magnitudes, even though symmetries are 
used. Nevertheless, $\Lambda_H$ is still considerably higher than the 
deviation of ground state energy. In particular, it exceeds 
$\left|\Delta E_0\right|$ by several orders of magnitudes for small $\lambda$.
Here we have to keep in mind that the truncation error does not only measure 
the inaccuracies of ground state energy, but the effect of truncation to the 
entire transformation of the Hamiltonian.

In contrast to the double hard-core boson model, no deviations in the maximal 
energy eigen-value were observed (not shown). This is explained
by the fact that the fully polarized state is still an exact eigen-state 
of the Hamiltonian \eqref{eq:spinhammi}.

For high values of $\lambda$, the truncation paradox occurs 
again because the truncation error bound $\widetilde\Lambda_H^{\dagger,xy}$ 
for the  three-triplon 
and the four-triplon truncation exceeds the truncation error bound of the 
two-triplon truncation. The exact truncation error $\Lambda_H$ 
does not display any paradoxical behaviour. Hence we conclude
that the extensive use of the triangle inequality is at the basis
of the truncation paradox.

\section{Summary\label{struct:facit}}

In this work, we presented a mathematically rigorous framework to bound
effects of truncation in self-similar continuous unitary transformations
\emph{a priori}. The difference ${H^{\prime\prime}}=H-H^{\prime}$ between 
a unitarily transformed Hamiltonian $H$ and the Hamiltonian
$H^{\prime}$ obtained from the truncated calculation is
captured by an inhomogeneous flow equation 
depending only on the truncated terms. We defined the scalar truncation error 
$\Lambda$ by the norm of the truncated terms 
$\left|\left|\kappa\right|\right|$. It provides an 
upper bound for the norm of the difference 
$\left|\left|H^{\prime\prime}\right|\right|$. 
A completely analogous bound is derived for observables as well.

Using the spectral norm, the truncation error implies an upper 
bound for the deviation of the minimal and maximal eigen-value of the 
observable under study, which is caused by truncation. 
In particular, we derived
a rigorous a priori bound for the error of ground state energy. 

The analysis of the double hard-core boson model showed that the norm of the 
difference ${H^{\prime\prime}}$ is bounded and approximated  very well by the 
truncation error. Despite the large difference to $\left|\Delta E_0\right|$, 
the truncation error $\Lambda_H$ provided a good measure for the deviation in 
the highest excited level $\left|\Delta E_\text{max}\right|$.
This could be understood by the special feature of the truncation scheme that 
primarily affected the highest excited level.

For practical use in extended systems, an upper bound $\widetilde\Lambda$ 
for the truncation error is calculated using the triangle inequality. 
The direct calculation of  $\left|\left|\kappa\right|\right|$ is not feasible 
-- even impossible for infinite systems -- because it would require an exact 
diagonalization in the Hilbert space of the entire system.

In both systems studied, the  double hard-core boson model and
the extended dimerized spin chain , the bound provided by 
truncation error $\Lambda_H$ turned out to overestimate the actual 
inaccuracies of the ground state energy significantly.
This is an inevitable consequence of the fact that the truncation error is a 
measure for the non-unitarity of the \emph{whole} transformation. Therefore 
it is sensitive to distortions of all eigen-values and eigen-states.

A comparison of various bounds for finite dimerized spin chain segments showed 
that the  truncation bound $\widetilde \Lambda_H$ overestimates the real 
truncation error $\Lambda_H$ by orders of magnitudes.
Furthermore, a truncation paradoxon was observed: Looser, i.e., better, 
truncation schemes implied higher truncation bounds. In the 
finite system studied using the ground state generator, 
this paradoxon did not occur for the exact truncation errors.

The truncation bound can be improved efficiently exploiting symmetries of the 
Hamiltonian and its hermitecity. By using the transposition of the triplon
polarizations $x$ and $y$ and the hermitecity, we were able to reduce the 
bound by a factor $\approx 2\sqrt{2}$. To overcome the high difference to 
the truncation error $\Lambda$, much more sophisticated approximations
would be needed. Additional symmetries available are  the cyclic spin 
permutation, the reflection symmetry, and the translation symmetry. 
We do not see a way to exploit the powerful translation symmetry
completely for the improvement of the bounds because this
requires finding bounds for operators in very large or infinite systems.
But larger subsets of terms of restricted range could indeed be analyzed
on larger clusters. The calculation of $H^{\prime\prime}$ in an extended 
system is left for further  investigation.

The analysis presented here provides a better understanding of 
truncation errors in continuous unitary transformations.
We are confident that it
serves as seed for stricter a priori error bounds in the future.

\begin{acknowledgments}
We are grateful for fruitful discussions with K.P.\ Schmidt in particular
for suggesting the illustrative model. We acknwoledge technical support by 
C.\ Raas. We want to thank S. Duffe for providing his CUT code 
at the initial stage of this work and for many helpful discussions.
\end{acknowledgments}

\appendix

\section{Effective system size\label{struct:app-effsize}}

\begin{figure}
  \includegraphics[width=\columnwidth]{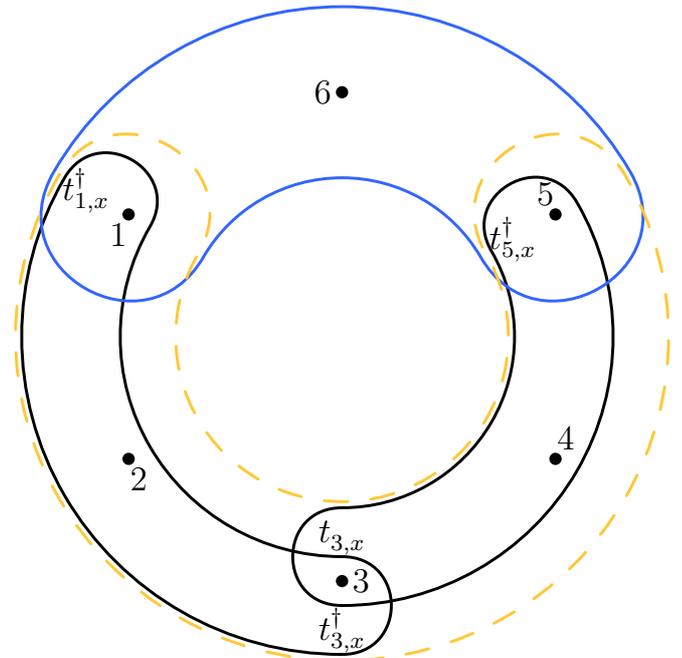}
  \caption{  (Color online) Schematic representation of the generation of a term
    which complies with the truncation scheme with $d_\text{max}=2$ only
    due to the finiteness of the systeme with size $l=6$. Following the 
    hard-core-algebra, the commutator of the terms 
    $t^\dagger_{1,x}t^\dagger_{3,x}$ and $t_{3,x}t^\dagger_{5,x}$ (black) 
    generates several terms, see Eq.\ \eqref{eq:A2}, with range 4 that are 
    truncated 
    (orange/gray and dashed), but the term $t_{1,x}t^\dagger_{5,x}$ is not
    truncated due  to a \glq wrap-around\grq\ effect (blue/gray).
    \label{img:effective}}
\end{figure}

\begin{table}
  \begin{tabular}{ccc}
    \hline
    $L$ 
    & $\lambda=0.5$ & $\lambda=1.0$  \\
    \hline
    3	&	-0.7821366540	&	-0.9342579563 \\
    4	&	-0.7778813863	&	-0.9019258071 \\
    5	&	-0.7770436419	&	-0.8836215559 \\
    6	&	-0.7768700368	&	-0.8745235739 \\
    7	&	-0.7768649893	&	-0.8740959679 \\
    8	&	-0.7768649893	&	-0.8740959679 \\
    9	&	-0.7768649893	&	-0.8740959679 \\
    10	&	-0.7768649893	&	-0.8740959679 \\
    \hline
  \end{tabular}
  \label{table:app-energies}
  \caption{Ground state energies per dimer $E_0$ for the finite dimerized 
    Heisenberg chain obtained by S-CUT using a real-space truncation scheme 
    for systems of different numbers of dimers $n$ and different interdimer 
    couplings $\lambda$. The maximum untruncated range is set to 
    $d_\text{max}=2$, the corresponding effective system size $L_\text{fin}$ 
    is 7. Beyond this size, 
    the calculated quantities become independent of the system size.}
\end{table}

In Sect.\ \ref{struct:math-effects} we mentioned that using a real-space 
truncation scheme implies an effective system size $L_\text{fin}$ that 
disguises the differences between infinite and finite periodic systems with 
a size of at least $L_\text{fin}$. The coefficients of representatives 
calculated by a truncated S-CUT become independent of the physical system size.
At first glance, this seems to be puzzling. But one has to look
in detail where the finite system size enters in the calculation of the flow 
equation.

For clarity, we focus on a one-dimensional system, although the argument can 
be generalized to higher dimensions. Let $d_\text{max}$ be the maximal range 
of the untruncated terms and $l$ the systems size. In order to keep the 
translational invariance of the infinite system also for a finite chain, 
periodic boundary conditions must be chosen. 
By evaluation of the flow equation \eqref{eq:flowequation}, the \emph{range} 
of the terms stemming from the commutator $\left[\eta(\ell),H(\ell)\right]$ 
becomes important.

For an infinite system, the issue is clear because the maximum
range of the commutator of two processes of range $d_1$ and $d_2$ is
given by the sum $d_\text{comm}=d_1+d_2$
$d_c=d_1+d_2$. In a periodic system
the range of a term is a more complex issue. 
The cluster of a term ambiguous because each site may be chosen to
be the leftmost site..
As an example, the range of the term $t^\dagger_{1,x}t^\dagger_{5,x}$ in a 
system of six dimers can be defined to be either 4 or 2, see Fig.\ 
\ref{img:effective}. To remove this ambiguity the most plausible prescription 
is to define the range of a term  to be the minimum range
of all possible choice of the leftmost site.

If the maximal range in the generator and in the Hamiltonian is $d_\text{max}$
the maximal range of terms stemming from the commutator is given by 
$d_\text{comm,max} = 2d_\text{max}$. Due to the locality of the algebra, the 
clusters of both terms have to share at least one common site as can be seen 
in Fig.\ \ref{img:effective}. So for smaller systems $L\le d_\text{comm,max}$
the result of the commutation is strongly influenced by the finite
size. If the system is larger than $d_\text{comm,max}$, the terms from the 
commutator are identical to those in the infinite system. But even then
a wrap-around can influence the range which is actually
attributed to a term resulting from the commutator.
As an example we consider the commutator 
\begin{align}
  \begin{split}
    & \left[t^\dagger_{1,x}t^\dagger_{3,x},t^{\phantom{\dagger}}_{3,x}
      t^\dagger_{5,x}\right]=
    -t^\dagger_{1,x}t^\dagger_{5,x}+2t^\dagger_{1,x}
    t^\dagger_{3,x}t^{\phantom{\dagger}}_{3,x}t^\dagger_{5,x}
    \\
    &\qquad
    +t^\dagger_{1,y}t^\dagger_{3,y}t^{\phantom{\dagger}}_{3,y}t^\dagger_{5,y}
    +t^\dagger_{1,z}t^\dagger_{3,z}t^{\phantom{\dagger}}_{3,z}t^\dagger_{5,z}.
    \label{eq:A2}
  \end{split}
\end{align}
In the infinite system, each of the terms on the right-hand side of 
\eqref{eq:A2} would be truncated since their range 4 exceeds the maximal range 
set to 2. But in the periodic system, the range of the contribution 
$-t^\dagger_{1,x}t^\dagger_{5,x}$ can be lower due to a 
\glq wrap-around\grq\ if the system is small enough, see Fig.\ 
\ref{img:effective}.

In general, this anomalous range assignment due to
a wrap-around can happen if the relation
\begin{align}
 L \leq d_\text{comm,max} + d_\text{max} = 3d_\text{max}
\end{align}
is fullfilled. As a result one obtains the same differential equations 
for the representatives for all systems  with a size 
of at least
\begin{align}
 L_\text{fin} = 3d_\text{max}+1.
 \label{eq:leff-def}
\end{align}
This included the infinite system. Hence $L_\text{fin}$
defines the effective size of a system treated with the maximal 
truncation range $d_\text{max}$.

For a numerical illustration, we examine the dependency of the ground state 
energy per dimer for the dimerized Heisenberg chain introduced in Sect.\ 
\ref{struct:extend}. We use the  real-space truncation scheme with a maximal 
truncation length $d_\text{max} =2$ and no restrictions of the number of 
interacting triplons. This implies an effective system size $L_\text{fin}$ of 
7 dimers according to \eqref{eq:leff-def}. The results are given in Tab.\ 
\ref{table:app-energies}. 
Clearly,  the results are numerically identical for larger systems
as predicted.

\bibliographystyle{apsrev}

\end{document}